\documentclass[%
 reprint,
superscriptaddress,
nofootinbib,
 amsmath,amssymb,
 aps,
prd,
]{revtex4-1}
\usepackage[T1]{fontenc}
\usepackage{aas_macros}
\usepackage{graphicx}
\usepackage{dcolumn}
\usepackage{bm}

\usepackage[citecolor=blue]{hyperref}
\usepackage[table]{xcolor}
\definecolor{medium-blue}{rgb}{0,0,1}
\hypersetup{colorlinks,linkcolor={magenta},citecolor={magenta},urlcolor={magenta}}  

\usepackage{ctable}
\usepackage{array}
\newcolumntype{C}[1]{>{\centering\arraybackslash\hspace{0pt}}p{#1}}
\newcolumntype{B}{>{\columncolor{blue!30}[.5\tabcolsep]}C{2.6cm}}
\begin{document}

\preprint{APS/123-QED}

\newcommand{\Hunit}{{\rm km}\ {\rm s}^{-1}\rm{Mpc}^{-1}}
\newcommand{\BB}[1]{\texttt{\color{red}[BB: #1]}}
\newcommand{\FK}[1]{\texttt{\color{blue}[FQ: #1]}}
\newcommand{\KS}[1]{\texttt{\color{green}[KS: #1]}}
\newcommand{\EC}[1]{\texttt{\color{orange}[EC: #1]}}
\newcommand{\jch}[1]{\texttt{\color{magenta}[JCH: #1]}}
\newcommand{\htj}[1]{\texttt{\color{yellow}[HTJ: #1]}}
\newcommand{\bds}[1]{\texttt{\color{blue}[BDS: #1]}}

\title{Accelerated inference on accelerated cosmic expansion:
New constraints on axion-like early dark energy with DESI BAO and ACT DR6 CMB lensing}

\author{Frank J. Qu}
\email{jq247@cantab.ac.uk}
\affiliation{DAMTP, Centre for Mathematical Sciences, Wilberforce Road, Cambridge CB3 0WA, UK}
\affiliation{Kavli Institute for Cosmology, University of Cambridge, Madingley Road, Cambridge CB3 0HA}

\author{Kristen M. Surrao}
\email{k.surrao@columbia.edu}
\affiliation{Department of Physics, Columbia University, New York, NY, USA 10027}

\author{Boris Bolliet}
\email{bb667@cam.ac.uk}
\affiliation{
Astrophysics Group, Cavendish Laboratory, J. J. Thomson Avenue, Cambridge CB3 0HE, United Kingdom
}
\affiliation{Kavli Institute for Cosmology, University of Cambridge, Madingley Road, Cambridge CB3 0HA}

\author{J.~Colin Hill}
\affiliation{Department of Physics, Columbia University, New York, NY, USA 10027}

\author{Blake D. Sherwin}
\affiliation{DAMTP, Centre for Mathematical Sciences, Wilberforce Road, Cambridge CB3 0WA, UK}
\affiliation{Kavli Institute for Cosmology, University of Cambridge, Madingley Road, Cambridge CB3 0HA}

\author{Hidde T. Jense}
\affiliation{School of Physics and Astronomy, Cardiff University, The Parade, Cardiff, Wales CF24 3AA, UK}

\date{\today}

\begin{abstract}
The early dark energy (EDE) extension to $\Lambda$CDM has been proposed as a candidate scenario to resolve the ``Hubble tension''.
We present new constraints on the EDE model by incorporating new data from the Dark Energy Spectroscopic Instrument (DESI) Baryon Acoustic Oscillation (BAO) survey and CMB lensing measurements from the Atacama Cosmology Telescope (ACT) DR6 and \textit{Planck} NPIPE data.  We do not find evidence for EDE. The maximum fractional contribution of EDE to the total energy density is $f_\mathrm{EDE}<  0.091 \; (95\% \; \mathrm{CL} )$ from our baseline combination of \textit{Planck} CMB, CMB lensing, and DESI BAO. Our strongest constraints on EDE come from the combination of \textit{Planck} CMB and CMB lensing alone, yielding $f_\mathrm{EDE}<  0.070 \; (95\% \; \mathrm{CL} )$. We also explore extensions of $\Lambda$CDM beyond the EDE parameters by treating the total neutrino mass as a free parameter, finding $\sum m_\nu < 0.096 \,\, {\rm eV} \; (95\% \; \mathrm{CL} )$ and $f_\mathrm{EDE}<  0.087 \; (95\% \; \mathrm{CL} )$.  For the first time in EDE analyses, we perform Bayesian parameter estimation using neural network emulators of cosmological observables, which are on the order of a hundred times faster than full Boltzmann solutions.
\end{abstract}

\maketitle


\section{\label{sec:level1}Introduction}

 One of the key parameters in cosmology, the Hubble constant $H_0$, has been determined with increasing precision from recent observational advances. On one hand, its value can be measured using indirect techniques, which depend on the assumption of a cosmological model. On the other hand, its value can be determined using direct local probes that are, to a large extent, free of these assumptions (with the caveat that it is assumed that these local probes are well-behaved at very low redshifts). The standard cosmological model, $\Lambda$ Cold Dark Matter ($\Lambda$CDM), predicts a value for $H_0$ based on observations of the cosmic microwave background (CMB), e.g., from \textit{Planck} \cite{Planck:2018_cosmo_params}, of $67.5\pm0.5\ \Hunit$. However, direct measurements of $H_0$ using Cepheid-calibrated Type Ia supernovae (SNIa) by the SH0ES collaboration \cite{Breuval:2024lsv} result in a higher $H_0=73.17\pm0.86\ \Hunit$, resulting in a $5\sigma$ tension with predictions based on $\Lambda$CDM, in what is commonly referred to as the ``Hubble tension''. Ref.~\cite{Verde:2019} discusses this tension in more detail, and Ref.~\cite{DiValentino:2021} reviews some of the attempts that have been made to resolve it. Other local $H_0$ measurements have included those from the Tip of the Red Giant Branch (TRGB), giving $H_0=69.8 \pm 0.6 \; \mathrm{stat} \pm 1.6 \; \mathrm{sys} \; \Hunit$ \cite{Freedman_2021} and the Hubble Space Telescope (HST) Key Project, giving $H_0=71 \pm 2 \; \mathrm{stat} \pm 6 \; \mathrm{sys} \; \Hunit $ \cite{2001ApJ...553...47F}. See Ref.~\cite{Freedman:2023} for a review on local, direct measurements.

Many attempts to resolve the Hubble tension involve scenarios beyond the $\Lambda$CDM model that increase the value of $H_0$ inferred from indirect probes. In this \emph{Letter}, we revisit the early dark energy (EDE) model using new data from the Dark Energy Spectroscopic Instrument (DESI) Baryon Acoustic Oscillation (BAO) survey and CMB lensing data from ACT and \emph{Planck} PR4. The EDE model (for reviews, see, e.g., \cite{McDonough:2023qcu,Poulin:2023lkg}) falls into the category of those that reduce the size of the sound horizon. In this model, a new field is introduced just before recombination to briefly accelerate the expansion relative to $\Lambda$CDM that decreases the sound horizon at recombination, consequently increasing $H_0$ in the fits to CMB data and alleviating the $H_0$ tension \cite{Poulin2019,Lin:2019qug,Agrawal:2019lmo,Kamionkowski:2022}.
We  focus on the axion-like EDE model, specified by the axion-like potential of the form~\cite{Poulin2019,Smith2019}
$$
V(\phi)=m^2f^2(1-\cos(\phi/f))^n .
$$
Following previous data analyses (e.g.,~\cite{Hill:2020osr,Ivanov2020,DAmico2020,LaPosta:2021pgm,PhysRevD.105.123536}), we restrict our analysis to integer $n=3$, with $n$ being the power-law index of the EDE potential and $m$ the mass of the field.

We parametrize this EDE model using effective parameters following the approach of, e.g.,~\cite{Poulin2019,Smith2019,PhysRevD.105.123536}. These parameters are given by the redshift $z_c$ at which EDE makes its largest fractional contribution $f_\mathrm{EDE}$ to the total cosmic energy budget, $$f_\mathrm{EDE}(z_c)=8\pi{G}\rho_\mathrm{EDE}(z_c)/(3H^2(z_c)) \, ,$$ and the initial field displacement $\theta_i\equiv\phi_i/f$, where $f$ is the decay constant.

Previous analyses usually include BAO as a probe of the low-redshift expansion rate with data from SDSS, BOSS, and eBOSS. BAO provide independent constraints on $\Omega_m$ to those from the CMB, which aid with degeneracy-breaking when including extensions like EDE. It is therefore worthwhile to investigate whether the constraints are robust to the BAO dataset used. This motivates the analysis with new BAO measurements from DESI-Y1 data, which have similar constraining power as BOSS/eBOSS. Similarly, CMB lensing is usually included as it mainly constrains the combination of $\sigma_8$ and $\Omega_m$ and hence further helps in breaking degeneracies compared to using the CMB alone.  Previous analyses used \textit{Planck} 2018 lensing measurements, but with the advent of new measurements from ACT DR6 and Planck PR4 with lower reconstruction noise, it is fitting to provide updated constraints on EDE.

Bayesian inference involving EDE models can be time-demanding: solving for the dynamics of the EDE field at a sufficiently high accuracy can take several minutes per step in a standard Markov Chain Monte Carlo (MCMC) chain and a typical computing set-up. Here we make use of neural network emulators constructed with \texttt{cosmopower} \cite{SpurioMancini:2021ppk}, following the same strategy as in \cite{Bolliet:2023sst}, to emulate the output of \texttt{CLASS\_EDE}~\cite{Hill:2020osr}.\footnote{\url{https://github.com/mwt5345/class_ede}} We incorporate our EDE emulators into \texttt{class\_sz} \cite{Bolliet:2023eob} so they can easily be used in Bayesian analysis with the \texttt{cobaya} sampler \cite{Torrado:2020dgo}, which we use throughout. Our machine-learning-accelerated pipeline allows us to reach convergence within $\mathcal{O}(10\;\mathrm{hours})$ instead of days or even weeks using full Boltzmann solutions. In all of our runs, we adopt a Gelman-Rubin convergence criterion with a threshold $R-1<0.01$. This improves upon previous EDE analyses that used, e.g., $R-1<0.05$ \cite{efstathiou2023improved}, $R-1<0.03$ \cite{PhysRevD.105.123536}, or $R-1<0.1$ \cite{Poulin2019}. We perform extensive checks of emulator accuracy in Appendix~\ref{app:b}, reproducing existing results on relevant datasets.

The priors adopted in this work are found in Appendix \ref{app.prior}.
Unless otherwise stated we use 3 degenerate massive neutrino states, and when the neutrino mass is fixed we use $\Sigma m_\nu = 0.06$ eV (with each neutrino carrying 0.02 eV). We keep the effective number of relativistic species at early times fixed to $N_\mathrm{eff}=3.046$. We work with a spatially flat $\Lambda$CDM(+EDE) cosmology throughout. We often refer to the derived parameters $S_8=\sigma_8(\Omega_m/0.3)^{0.5}$ and $h=H_0/(100\,\mathrm{km\,s^{-1}\,Mpc^{-1}})$.

\section{Data and likelihoods}

In this analysis, we focus on CMB data and the improvements in constraints that the new DESI BAO and ACT DR6 CMB lensing data add. The datasets used are detailed below.

\textbf{\textit{Planck} CMB:} We include temperature and polarization power spectra of the primary CMB as observed by the \textit{Planck} satellite. Specifically, we use the small scale $\ell>30$ TT, TE, and EE bandpowers analyzed from the \textit{Planck} Public Release 4 (PR4) (NPIPE, \cite{Planck:2020olo}) maps based on the \texttt{Camspec} likelihood \cite{Rosenberg:2022sdy}. We also include the \textit{Planck} PR3 likelihood for the large-scale temperature power spectrum and large-scale polarization information that constrains the optical depth to reionization using the likelihood from the \texttt{Sroll} maps \cite{pagano/etal:2020}. We will subsequently refer to the above combination as \textit{Planck} CMB. We note that for one of our benchmark runs we also use \textit{Planck} PR3 TTTEEE data, see Appendix \ref{app.planck19benchmark}.

 \textbf{CMB lensing:} We employ CMB lensing power spectrum data from the Atacama Cosmology Telescope (ACT) DR6 \cite{ACT:2023dou,ACT:2023kun,ACT:2023ubw} and \textit{Planck} PR4 (NPIPE) \cite{Planck:2020olo}. The ACT DR6 lensing map covers $9400,\rm{deg}^2$ and is signal dominated on scales $L < 150$, achieving a precision of $2.3\%$ ($43\sigma$). The \textit{Planck} PR4 lensing analysis benefits from reprocessed maps with around $8\%$ more data than PR3, resulting in a $20\%$ increase in signal-to-noise ratio compared to the 2019 \textit{Planck} PR3 release. We refer to the combined lensing likelihood from both experiments as our baseline, denoted as CMB lensing. We write ``CMB lensing 2018'' when we use the \textit{Planck} 2018 CMB lensing likelihood ~\cite{planck2018lensing}.

\textbf{DESI BAO:} We consider BAO measurements from the DESI-Y1 release ~\cite{desicollaboration2024desi}. DESI measured BAO from the clustering of galaxies with samples spanning redshifts $0.1\leq{z}\leq4.2$. These include seven redshift bins comprising bright galaxy samples (BGS), luminous red galaxies (LRG), emission line galaxies (ELG), quasars (QSO) and the Lyman-$\alpha$ forest sample. We use the official DESI likelihood, publicly available in Cobaya. In Appendix \ref{app.desi} we show that we can recover the constraints in \cite{desicollaboration2024desi}. We denote the above as DESI BAO.


\textbf{Pre-DESI BAO (and RSD):} When specified, we also test data combinations that utilize BAO and redshift-space distortion (RSD) measurements from 6dFGS \cite{2011MNRAS.416.3017B}, the SDSS DR7 Main Galaxy Sample (MGS;~\cite{1409.3242}), BOSS DR12 luminous red galaxies (LRGs; \cite{1607.03155}), and eBOSS DR16 LRGs \cite{2007.08991}, which we will subsequently denote as pre-DESI BAO (pre-DESI BAO and RSD when including growth information from RSD).

\textbf{Pantheon+}: In certain data combinations we make use of SNIa from the Pantheon+ compilation \cite{Scolnic_2022, Brout:2022vxf}, which comprises 1550 spectroscopically confirmed SNIa in the redshift range $0.001\leq{z}\leq2.26$.

\textbf{SH0ES}: This refers to the addition of SH0ES Cepheid host distance anchors to Pantheon+ (see \cite{Brout:2022vxf} for details). This is similar to adding a Gaussian prior on the peak SN1a absolute magnitude, $M_b$ as in \cite{efstathiou2023improved}, or a Gaussian prior on $H_0$ as in, e.g., \cite{PhysRevD.105.123536}, but without approximation.

\begin{figure*}
\includegraphics[width=\textwidth]{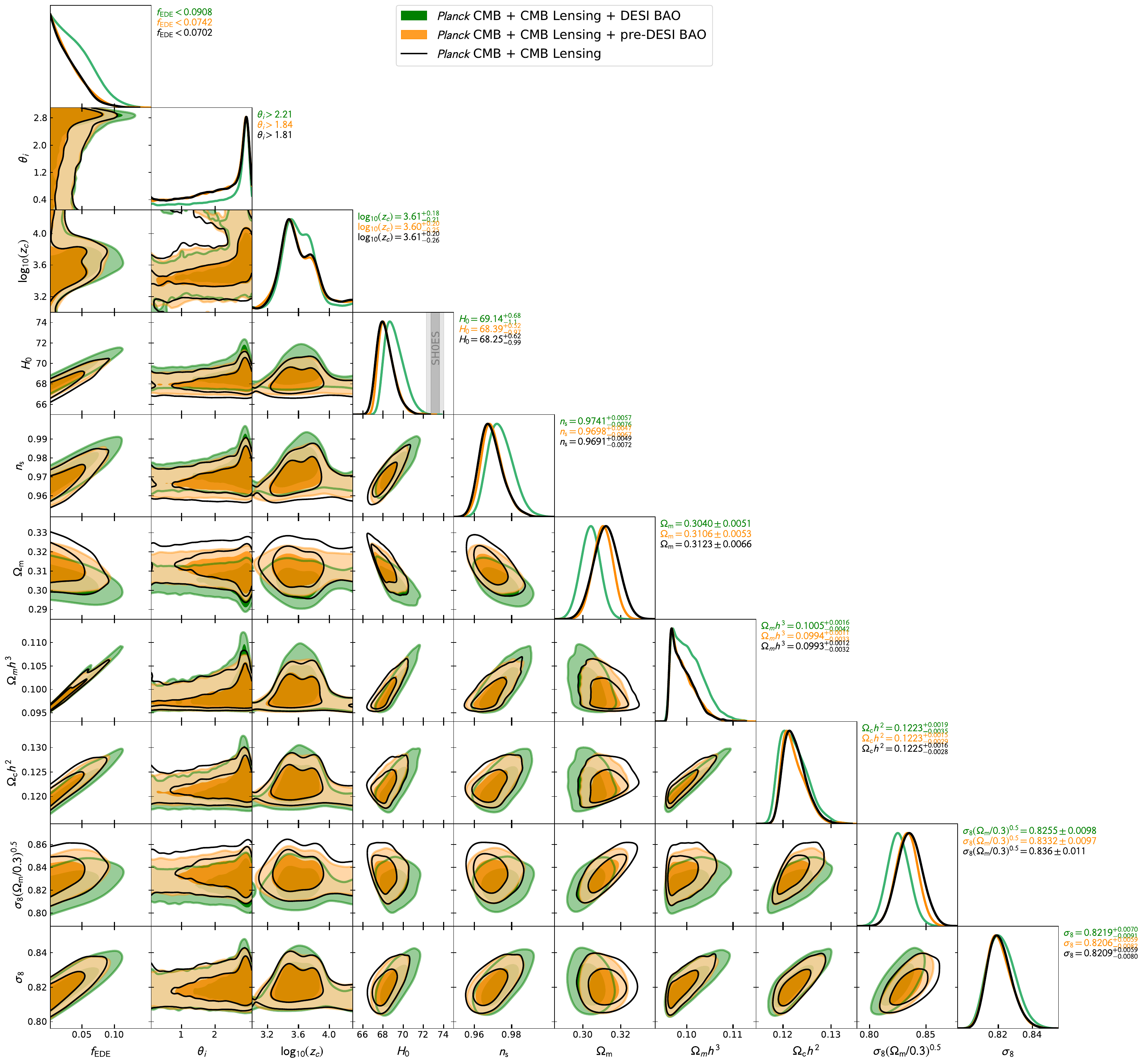}
\caption{Marginalized (1D and 2D) joint posterior probability distributions for the EDE parameters and a subset of other parameters in fits to our baseline \textit{Planck} CMB + CMB lensing + DESI BAO data (green), \textit{Planck} CMB + CMB lensing + pre-DESI BAO data (orange), and \textit{Planck} CMB + CMB lensing only (black). The vertical grey bands in the $H_0$ panel show the latest SH0ES constraint \cite{Breuval:2024lsv} as a reference. }
\label{fig:EDE_MAIN}
\end{figure*}

\section{Results}

\subsection{New EDE constraints with \textit{Planck} CMB + CMB lensing + DESI BAO}\label{res.1}

  \begin{table*}[ht!]
Constraints on EDE ($n=3$)  \\
  \centering
  \begin{tabular}{|c|c|c|c|c|c|c|}
    \hline\hline Parameter &  \hspace{0cm}\begin{tabular}[t]{@{}c@{}}\textbf{\textit{Planck}} \textbf{CMB}\\\textbf{CMB Lensing} \\ \textbf{DESI BAO}\end{tabular} 
    & \hspace{0cm}\begin{tabular}[t]{@{}c@{}}{\textit{Planck}} {CMB}\\CMB Lensing \\ pre-DESI BAO\end{tabular}
    &\hspace{0cm}\begin{tabular}[t]{@{}c@{}}{\textit{Planck}} {CMB}
    \\CMB Lensing \end{tabular} 
    &\hspace{0cm}\begin{tabular}[t]{@{}c@{}}{\textit{Planck}} {CMB}\\DESI BAO\end{tabular}
    & \hspace{0cm}\begin{tabular}[t]{@{}c@{}}{\textit{Planck}} {CMB}\end{tabular}
    &\hspace{0cm}\begin{tabular}[t]{@{}c@{}}\emph{Planck} 2018 \\ TT+TE+EE \\\end{tabular} 
    \\\hline \hline

    {\boldmath$f_\mathrm{EDE} $}   
    &$<0.091$  
    &$ < 0.074 $ 
    &$ < 0.070 $
    &$ < 0.093 $ 
    &$ < 0.081 $
    &$ < 0.093 $  \\

    {\boldmath$\mathrm{log}_{10}(z_c)$}   
    & $3.61^{+0.18}_{-0.21}$ 
    & $3.60^{+0.20}_{-0.25}$ 
    & $3.61^{+0.20}_{-0.26}$
    &$3.63^{+0.19}_{-0.22}$ 
    & $3.61^{+0.21}_{-0.24}$
    & $3.63\pm 0.24$ \\


    \hline
    
    $H_0 \, [\mathrm{km/s/Mpc}]$   
    & $69.14^{+0.68}_{-1.10}$ 
    &$68.39^{+0.52}_{-0.97}$
    & $68.25^{+0.62}_{-0.99}$
    &$69.19^{+0.66}_{-1.20}$
    &$68.37^{+0.68}_{-1.10}$
    &$68.42^{+0.79}_{-1.30}$ \\

    $\Omega_m$   
    &$0.3027^{+0.0064}_{-0.0038} $  
    & $0.3093^{+0.0066}_{-0.0038} $ 
    & $0.3109^{+0.0080}_{-0.0052} $
    & $0.3016^{+0.0067}_{-0.0040} $
    & $0.3106\pm 0.0088 $
    & $0.3146\pm 0.0088$
    \\

    $\Omega_ch^2$   
    &$0.1223^{+0.0019}_{-0.0035}$  
    & $0.1223^{+0.0015}_{-0.0029}$ 
    & $0.1225^{+0.0016}_{-0.0028}$
    & $0.1220^{+0.0019}_{-0.0038}$
    & $0.1228^{+0.0018}_{-0.0033}$
    & $ 0.1240^{+0.0021}_{-0.0041}$
    \\

    $S_8$   
    & $0.826\pm{0.010}$
    & $0.833\pm{0.010}$
    & $0.836\pm{0.011}$
    & $0.821\pm{0.012}$
    & $0.835\pm{0.015}$  
    & $0.843\pm{0.018}$\\
    
    $\sigma_8$   
    &$0.822^{+0.007}_{-0.009}$ 
    & $0.821^{+0.006}_{-0.008}$ 
    & $0.821^{+0.006}_{-0.008}$
    & $0.818^{+0.009}_{-0.011}$
    & $0.821^{+0.008}_{-0.011}$ 
    & $0.823^{+0.010}_{-0.013}$\\

    \hline

    $R-1$
    &$0.006981$ 
    & $0.004575$ 
    & $0.009128$
    & $0.006200$
    & $0.009568$ 
    & $0.000990$\\

    \hline
  \end{tabular} 
  \caption{ Marginalized constraints on key cosmological parameters in the EDE model (with power-law index $n=3$) in bold and other relevant cosmological parameters. Upper limits are given at the $95\%$ CL while constraints are given at $68\%$ CL.}

  \label{table:EDE-params-full} 
\end{table*}

The baseline data combination adopted in this work with \textit{Planck} CMB, new CMB lensing from ACT DR6 and \textit{Planck} PR4 NPIPE, and new DESI BAO Y1 provides the following upper bound on EDE
\begin{equation}
    f_\mathrm{EDE}< 0.091 \, (0.0512) \quad   95\% \, (68\%)  \mathrm{CL}.
\end{equation}
The marginalized posterior probability distribution for this analysis is shown in Fig.~\ref{fig:EDE_MAIN} as the green contours.  For the Hubble constant we find
\begin{equation}
H_0 = 69.14^{+0.68}_{-1.1} \,\, \Hunit.
\end{equation}
This is in $4.7\sigma$ tension with the latest SH0ES-inferred value of $73.17\pm0.86 \,\, \Hunit$ \cite{Breuval:2024lsv} (see grey vertical band in the $H_0$ panel of Fig.~\ref{fig:EDE_MAIN}).

For completeness, we carry out the same analysis, substituting DESI BAO with pre-DESI BAO (orange contours in Fig.~\ref{fig:EDE_MAIN}). In this case we find a slightly tighter bound on EDE, namely: 
\begin{equation}
    f_\mathrm{EDE}< 0.074 \;  (0.036) \;   95\% (68\%) \\ \mathrm{CL}.
\end{equation}
The corresponding $H_0$ constraint is: 
\begin{equation}
H_0 = 68.39^{+0.52}_{-0.97} \,\, \Hunit,
\end{equation}
in $5.6\sigma$ tension with the latest SH0ES-inferred value.

In Fig.~\ref{fig:EDE_MAIN}, we also show the resulting constraints for the same analysis without BAO data in black. This combination of \textit{Planck} CMB and CMB lensing, i.e., ACT DR6 + \textit{Planck} PR4 (NPIPE), yields our tightest EDE bound: 
\begin{equation}
    f_\mathrm{EDE}< 0.070 \; (0.036) \quad   95\% \,(68\%) \  \mathrm{CL}.
\end{equation}

Comparing contours in Fig.~\ref{fig:EDE_MAIN}, we see that the main effect of adding pre-DESI BAO to \textit{Planck} CMB and CMB lensing is a tightening of the $\Omega_m$ constraint. The effects on other parameters, including EDE parameters, are marginal. 

Nonetheless, adding DESI BAO to \textit{Planck} CMB and CMB lensing has an appreciable effect not only on $\Omega_m$\footnote{We note that the constraining power of pre-DESI and DESI BAO on $\Omega_m$ is nearly the same, but DESI BAO prefers a lower $\Omega_m$.} but also on $n_s$, $H_0$, and $f_\mathrm{EDE}$. The weakening of the upper bound on $f_\mathrm{EDE}$ (see top frame in triangle plot of Fig.~\ref{fig:EDE_MAIN}) can be attributed to the fact that DESI BAO data is pushing the model towards a matter fraction that is 1.2$\sigma$ lower than the $\Omega_m$ value preferred by pre-DESI BAO and CMB data. Indeed, the combination of CMB and BAO data leads to a negative degeneracy between $\Omega_m$ and $H_0$, as well as $\Omega_m$ and $n_s$, while both $H_0$ and $n_s$ have a positive degeneracy with $f_\mathrm{EDE}$.

In Table \ref{table:EDE-params-full} we report constraints on a relevant subset of parameters for the analyses of Fig.~\ref{fig:EDE_MAIN}, as well as for other combinations of datasets: without lensing, with \textit{Planck} CMB only (i.e., PR4) and with \textit{Planck} 2018 CMB (i.e., PR3). We find that switching from \textit{Planck} PR3 to \textit{Planck} PR4 CMB yields a $\approx 10\%$ more constraining bound on $f_\mathrm{EDE}$. As discussed above, adding DESI BAO slightly degrades the bound because of the lower $\Omega_m$. Moreover, comparing the results in the third and fifth column show that adding CMB lensing tightens the bound on  $f_\mathrm{EDE}$ by another $\approx 10\%$. Marginalized posterior probability distributions for all analyses in Table \ref{table:EDE-params-full} (except the last column with \textit{Planck} 2018) are shown in Fig.~\ref{fig:full_contours} in  Appendix \ref{app:fullcontours}.

\begin{figure*}
\includegraphics[width=\textwidth]{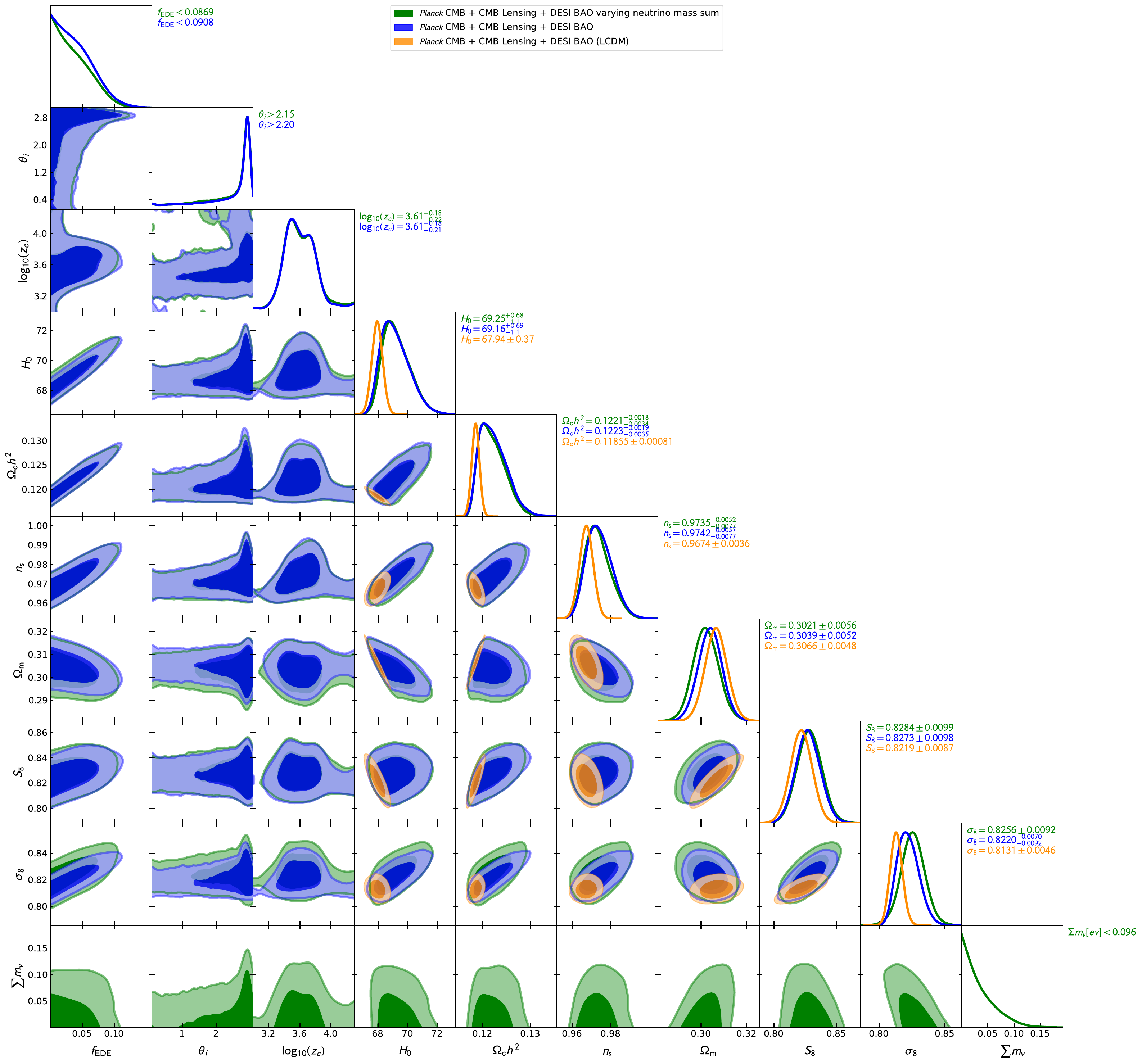}
\caption{Marginalized (1D and 2D) joint posterior probability distributions for the EDE parameters and a subset of other parameters in fits to our baseline \textit{Planck} PR4 (NPIPE) TT+TE+EE data + ACT DR6 and PR4 CMB lensing + DESI BAO data (blue) compared to the analysis obtained with the same data combination but allowing the neutrino mass to vary (green). In orange we show the same dataset analyzed within $\Lambda$CDM.
}
\label{fig:mnu}
\end{figure*}

Making use of our Boltzmann code emulators, we can study further extensions to the $\Lambda$CDM model. In particular, we can investigate the stability of the neutrino mass bound obtained in \citep{desicollaboration2024desi} in cosmological models with EDE. Using our baseline dataset (\textit{Planck} CMB + CMB lensing + DESI BAO), we consider a model with EDE and massive neutrinos, with the  neutrino mass sum as a free parameter. The contours are shown in green in Fig.~\ref{fig:mnu} together with the fixed neutrino mass analysis (blue contours) and $\Lambda$CDM as a reference in orange. In agreement with \cite{Reeves_2023}, we find no statistically significant degeneracies between the EDE parameter space and $\sum{m}_\nu$.  The addition of free neutrino mass does not change the constraint on $f_\mathrm{EDE}$ in a substantial way. Conversely, adding EDE changes the neutrino mass bound by only a small amount, going from  $\sum{m}_\nu<0.072 \,\, {\rm eV} \, (95\% \, \mathrm{CL})$ in $\Lambda$CDM \citep{desicollaboration2024desi} to $\sum{m}_\nu<0.097 \,\, {\rm eV} \, (95\% \, \mathrm{CL})$.

Given that EDE parameters and $\sum{m}_\nu$ do not have statistically significant degeneracies, we only report results keeping the neutrino mass sum fixed to the minimum value allowed by the normal hierarchy of $\sum{m_\nu}=0.06$ eV in the remainder of this paper. 

\subsection{Inclusion of Pantheon+ SNIa} \label{sec.pantheon}

In \cite{efstathiou2023improved}, upper bounds on $f_\mathrm{EDE}$ were provided using a combination of
\textit{Planck} PR4 (NPIPE) TT/TE/EE, low-$\ell$ TT and EE likelihoods\footnote{The CMB primary combination used here is similar to that of our baseline analysis, with the exception of the low-$\ell$ EE data, where we use a more updated version based on {\tt Sroll}; we will thus denote this CMB primary combination as CMB (low $\ell$ EE 2018).} from \textit{Planck} 2018, \textit{Planck} 2018 CMB lensing, measurements of the BAO and RSD from the CMASS and LOWZ galaxy samples of BOSS DR12, 6dFGS, and SDSS DR7, and the Pantheon+ catalogue of over 1600 SNIa.

For the same data combination, but setting a more stringent convergence criterion of $R-1<0.01$ (compared to the $R-1<0.05$ of \cite{efstathiou2023improved}), we find 
\begin{align}
    f_\mathrm{EDE} < 0.070 \; (95\% \mathrm{CL}),\,\, 
    h=0.6814^{+0.0050}_{-0.0092} \;(68\% \mathrm{CL}). 
\end{align}
(See the first column at the top of Table~\ref{table:george_comparison.} for details.) This is consistent with the results from \cite{efstathiou2023improved}, namely, 
\begin{equation}
f_\mathrm{EDE}<0.061 \;(95\% \mathrm{CL})\,\, h=0.6811^{+0.0048}_{-0.0082}\;(68\% \mathrm{CL}) .
\end{equation}

Furthermore, we consider the addition of the SH0ES Cepheid host distances to Pantheon+, similar to the Gaussian prior on $M_b$ used in \cite{efstathiou2023improved}, but without approximation. 

Our results are in the second  column at the top of Table~\ref{table:george_comparison.}. We find 
\begin{align}
    f_\mathrm{EDE} &= 0.110\pm 0.023 \; (68\% \mathrm{CL}),\,\,\\ 
    h &=0.7120\pm 0.0079 \;(68\% \mathrm{CL}), 
\end{align}
consistent with the results from \cite{efstathiou2023improved}, namely, 
\begin{align}
    f_\mathrm{EDE} &= 0.107\pm 0.023 \; (68\% \mathrm{CL}),\,\,\\ 
    h&=0.7124\pm 0.0077 \;(68\% \mathrm{CL}). 
\end{align}

The small differences between our results and those of \cite{efstathiou2023improved} can be attributed to a combination of the different convergence of the chains\footnote{By truncating the initial and final parts of our converged chains, we checked that larger $R-1$ values are associated with more variance in the estimated bounds, typically $10\%$ for $R-1\approx 0.05$.} and possibly different implementations of the EDE model.\footnote{We use the public \texttt{class\_ede} code~\cite{Hill:2020osr}, while \cite{efstathiou2023improved} refers to a different modified version of \texttt{class}.  }

\begin{figure*}
\includegraphics[width=\textwidth]{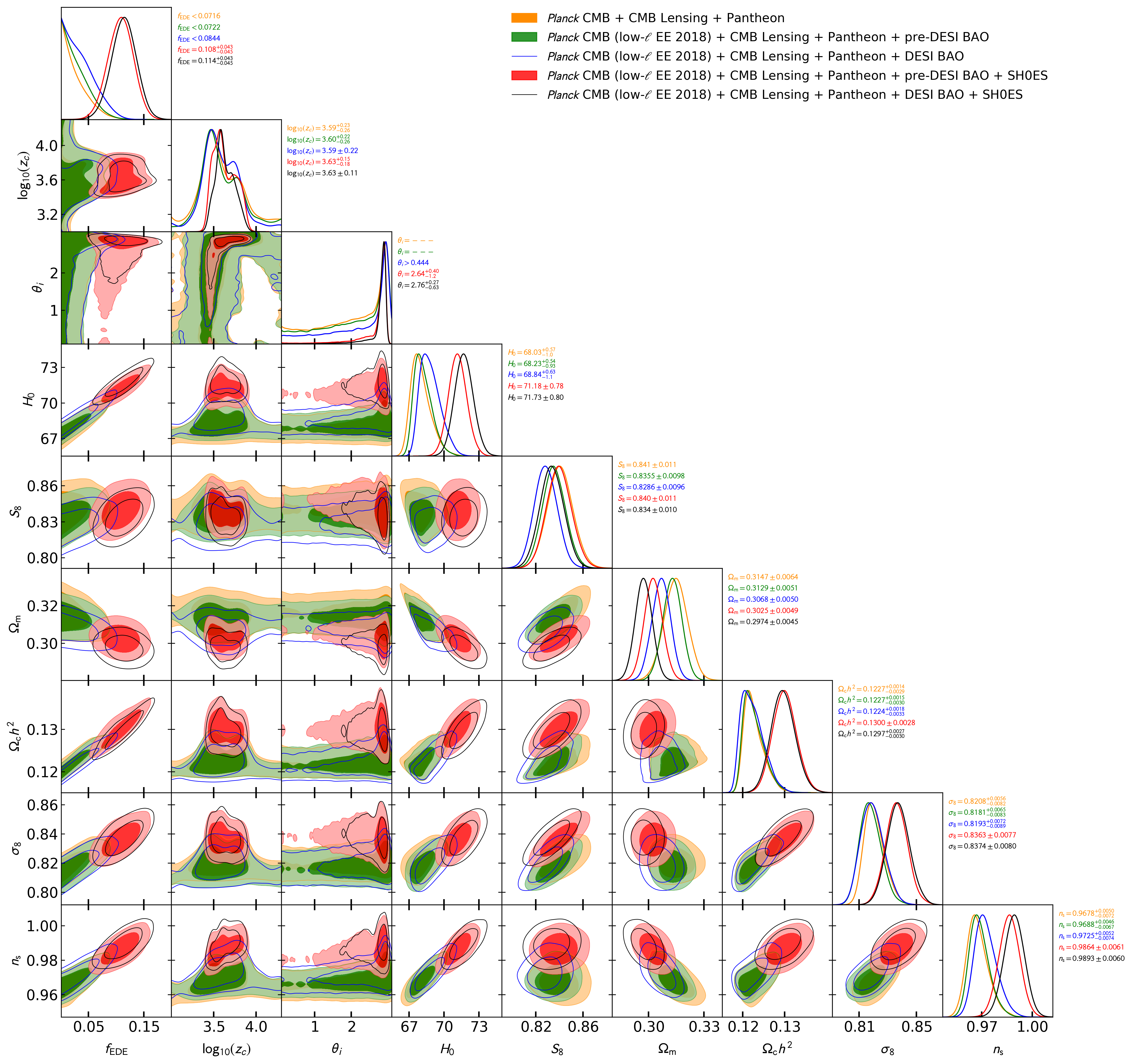}
\caption{Marginalized (1D and 2D) joint posterior probability distributions for the EDE parameters and a subset of other cosmological parameters of interest in analyses that include the Pantheon+ Type Ia supernova dataset, with and without SH0ES Cepheid host distance anchors. See Sec.~\ref{sec.pantheon} for details.}
\label{fig:george}
\end{figure*}

In Table~\ref{table:george_comparison.}, we provide updated versions of these bounds using the same dataset but replacing the CMB lensing and BAO data with the new  ACT DR6 and \textit{Planck} PR4 (NPIPE) CMB lensing measurement and DESI BAO data. In Fig.~\ref{fig:george} we show the marginalized posterior probability distributions for the updated constraints. We also plot the contours (orange) for an analysis without BAO but including Pantheon+ (as well as CMB and CMB lensing), yielding $f_\mathrm{EDE}<0.0716$ (95\%CL). Similar to the previous section, we find that the tightest bound on EDE is achieved in the analysis without BAO.

  \begin{table*}
Constraints on EDE ($n=3$)  \\
  \centering
  \begin{tabular}{|c|c|c|c|c|}
    \hline\hline Parameter &  \hspace{0cm}\begin{tabular}[t]{@{}c@{}} CMB (low $\ell$ EE 2018) \\
    CMB Lensing 2018 \\
    pre-DESI BAO+RSD \\
    Pantheon+\end{tabular} 
    & \hspace{0cm}\begin{tabular}[t]{@{}c@{}}CMB (low $\ell$ EE 2018) \\
    CMB Lensing 2018 \\
    pre-DESI BAO+RSD \\
    Pantheon+\\
    SH0ES $M_b$ prior \\ \end{tabular}
    &\hspace{0cm}\begin{tabular}[t]{@{}c@{}}CMB (low $\ell$ EE 2018) \\
    CMB Lensing 2018 \\
    DESI BAO \\
    Pantheon+
    \\ \end{tabular}
    &\hspace{0cm}\begin{tabular}[t]{@{}c@{}}CMB (low $\ell$ EE 2018) \\
    CMB Lensing 2018 \\
    DESI BAO\\
    Pantheon+\\
    SH0ES $M_b$ prior \end{tabular} 
    \\\hline \hline

    {\boldmath$f_\mathrm{EDE} $}   
    &$< 0.070$  
    &$ 0.110\pm 0.023$ 
    &$ < 0.087 $
    &$ 0.114\pm 0.023$  \\

    {\boldmath$\mathrm{log}_{10}(z_c)$}   
    & $3.60^{+0.23}_{-0.26}$ 
    & $3.64^{+0.17}_{-0.19}$ 
    & $ 3.61^{+0.20}_{-0.23} $
    & $3.64\pm 0.11$ \\

    \hline
    
    $H_0 \, [\mathrm{km/s/Mpc}]$   
    & $68.14^{+0.50}_{-0.92}$ 
    &$71.20\pm 0.79 $
    &$68.83^{+0.60}_{-1.1}$
    &$71.69\pm 0.80$ \\

    $\Omega_m$   
    & $0.3130\pm 0.0052$ 
    &$0.3031\pm 0.0050$
    &$0.3067\pm 0.0051$
    &$0.2980\pm 0.0046$ \\

    $\Omega_ch^2$   
    & $0.1224^{+0.0013}_{-0.0029}$
    & $0.1304\pm 0.0030$
    &$0.1223^{+0.0017}_{-0.0035}$
    &$0.1299\pm 0.0029$ \\

    $S_8$   
    & $0.834\pm 0.011$ 
    &$0.842\pm 0.012 $
    &$0.827\pm 0.011 $
    &$0.835\pm 0.011$ \\
    
    $\sigma_8$   
    & $0.8165^{+0.0070}_{-0.0091}$ 
    &$ 0.8375\pm 0.0092$
    &$ 0.8179^{+0.0080}_{-0.010}$
    &$0.8378\pm 0.0091$ \\
    \hline

    $R-1$   
    & $0.00283$ 
    &$0.00866$
    &$0.00306$
    &$0.00674$ \\

    \hline
  \end{tabular}

  \begin{tabular}{|c|c|c|c|c|}
    \hline\hline Parameter
    &\hspace{0cm}\begin{tabular}[t]{@{}c@{}}CMB (low $\ell$ EE 2018) \\
    CMB Lensing \\
    pre-DESI BAO+RSD \\
    Pantheon+\end{tabular}
    & \hspace{0cm}\begin{tabular}[t]{@{}c@{}}CMB (low $\ell$ EE 2018) \\
    CMB Lensing \\
    pre-DESI BAO+RSD \\
    Pantheon+\\
    SH0ES $M_b$ prior 
    \end{tabular}
    &\hspace{0cm}\begin{tabular}[t]{@{}c@{}}CMB (low $\ell$ EE 2018) \\
    CMB Lensing \\
    DESI BAO \\
    Pantheon+\\ \\\end{tabular} 
     &\hspace{0cm}\begin{tabular}[t]{@{}c@{}}
    CMB (low $\ell$ EE 2018) \\
    CMB Lensing \\
    DESI BAO \\
    Pantheon+\\
    SH0ES $M_b$ prior 
    \end{tabular} 
    \\\hline \hline

    {\boldmath$f_\mathrm{EDE} $}   
    & $<0.073$ 
    &$0.108\pm 0.023$
    &$<0.080$
    &$0.114\pm 0.022$ \\

    {\boldmath$\mathrm{log}_{10}(z_c)$}   
    & $3.60^{+0.21}_{-0.26}$ 
    &$3.63^{+0.15}_{-0.18}$
    &$3.59^{+0.19}_{-0.21}$ 
    &$3.63\pm 0.11$ \\

    \hline
    
    $H_0 \, [\mathrm{km/s/Mpc}]$   
    & $68.23^{+0.54}_{-0.93}$ 
    &$71.18\pm 0.78$
    &$68.13^{+0.48}_{-0.86}$ 
    &$71.73\pm 0.80$ \\

    $\Omega_m$   
    & $0.3115^{+0.0065}_{-0.0037}$ 
    &$0.3025\pm 0.0049$
    &$0.3055^{+0.0063}_{-0.0037}$ 
    &$0.2974\pm 0.0045$ \\

    $\Omega_ch^2$   
    & $0.1227^{+0.0015}_{-0.0030}$
    & $0.1300\pm 0.0028$
    & $0.1224^{+0.0018}_{-0.0034}$
    &$0.1297^{+0.0027}_{-0.0030}$ \\

    $S_8$   
    & $0.8336\pm{0.0098}$ 
    &$0.840\pm 0.011$
    &$0.8268\pm{0.0096}$ 
    &$0.834\pm 0.010$ \\
    
    $\sigma_8$   
    & $0.8181^{+0.0065}_{-0.0082}$ 
    &$0.8363\pm 0.0077$
    &$0.8194^{+0.0071}_{-0.0089}$ 
    &$0.8374\pm 0.0080$ \\
    
    \hline

    $R-1$   
    & $0.00960$ 
    &$0.00918$
    &$0.00697$
    &$0.00566$ \\

    \hline
  \end{tabular} 
  
  \caption{Marginalized constraints on key cosmological parameters in the EDE model (with power-law index $n=3$).  EDE parameters are in bold. Upper limits are given at the $95\%$ CL while constraints are given in $68\%$ CL. The first two column of the first row shows constraints obtained with a data combination that matches that of \cite{efstathiou2023improved}. We then explore the effects of including new DESI BAO and CMB lensing from ACT DR6 and \textit{Planck} PR4 (NPIPE).}
  \label{table:george_comparison.}
\end{table*}

\section{Discussion and Conclusion}

\begin{table*}
\centering
\begin{tabular}{|c|c|c|c|}
\hline\hline 
 &  $\Lambda$CDM & EDE & $\Delta\chi^2=\chi^2_\mathrm{EDE}-\chi^2_{\Lambda\mathrm{CDM}}$ \\ \hline \hline
\texttt{Camspec} NPIPE TTTEEE & $10542.61$ & $10542.60$ &  $-0.01$  \\ \hline
CMB Lensing &  $19.81$ & $20.38$  & $0.57$ \\ \hline
DESI BAO & $15.64$  &  $14.01$ & $-1.63$  \\ \hline
$\chi^2_\mathrm{total}$&  $10578.06$ &  $10576.99$&  $-1.07$\\ \hline

\end{tabular}
  \caption{Best fit $\chi^2$ for each experiment and the total $\chi^2_\mathrm{total}$.  These values are reported for the $\Lambda\mathrm{CDM}$ and EDE models with our baseline data set. The $\Delta\chi^2$ values for the $\Lambda\mathrm{CDM}$ and EDE models are also shown.}
\label{table:chi2_1}
\end{table*}

\begin{table*}
\centering
\begin{tabular}{|c|c|c|c|c|}
\hline\hline 
 &  $\Lambda$CDM & EDE & EDE with Pantheon+ & EDE with Pantheon+ + SH0ES prior\\ \hline \hline
\texttt{Camspec} NPIPE TTTEEE & $10542.6$ & $10542.6$ &  $10541.1$ &  $10546.0$ \\ \hline
CMB Lensing &  $19.8$ & $20.4$  & $20.6$ &  $21.0$\\ \hline
DESI BAO & $15.6$  &  $14.0$ & $14.9$&  $12.8$  \\ \hline
Pantheon+&-  & -&  $1405.3$&  $1460.5$\\ \hline
$\chi^2_\mathrm{total}$&  $10578.0$ &  $10577.0$&  $11981.9$ &$12040.3
$\\ \hline
$\chi^2_\mathrm{com}$&  $10578.0$ &  $10577.0$&  $10576.6$ &$10579.8$\\ \hline
$\chi^2_\mathrm{com, no DESI}$&  $10562.4$ &  $10563.0$&  $10561.7$ &$10567.0$\\ \hline
\end{tabular}
    \caption{Best fit $\chi^2$ for each experiment, the total $\chi^2_\mathrm{total}$, and 
 $\chi^2_\mathrm{com}$ which consists of the subset of experiments common to all the combinations reported, specifically \texttt{Camspec} NPIPE TTTEEE, CMB lensing, and DESI BAO. These values are reported for the $\Lambda\mathrm{CDM}$ and EDE models with our baseline data set. Furthermore, $\chi^2_\mathrm{com,noDESI}$ is the same as $\chi^2_\mathrm{com}$ but excludes DESI BAO. }
 \label{table:chi2_2}
\end{table*}

This \textit{Letter} provides updated constraints on axion-like EDE in light of new BAO data from DESI Y1 and CMB lensing measurements from ACT DR6 and \textit{Planck} PR4. 

Our main results are summarized in Table \ref{table:EDE-params-full} and Table \ref{table:george_comparison.}. We find that using CMB and CMB lensing alone, one can place strong constraints on the maximum fractional contribution of EDE to the total energy density, with  $f_\mathrm{EDE} < 0.070 \: (95\%)$. 
The addition of DESI slightly degrades this bound to $f_\mathrm{EDE}<  0.091 \; (95\% \; \mathrm{CL} )$ due to the low value of $\Omega_m$ preferred by DESI. Nevertheless the data do not show any statistically significant preference for EDE. This is shown in the black open contour of our main plot in Fig.~\ref{fig:EDE_MAIN}.  As a guide, it is pointed out in ~\cite{Poulin2019,Smith2019,PhysRevD.105.123536}  that a $f_\mathrm{EDE}=0.1$ at a redshift $z_c$ around matter-radiation equality is required for EDE to be a viable model in resolving the Hubble tension. 

The lack of EDE preference is confirmed by comparing $\chi^2$ values of the various experiments for $\Lambda$CDM to those from EDE, as shown in Tables \ref{table:chi2_1} and \ref{table:chi2_2}. When adding the three additional EDE parameters, there is a total $\Delta \chi^2 \equiv \chi^2_{\mathrm{EDE}} - \chi^2_{\Lambda \mathrm{CDM}}=-1.07$ (including \texttt{Camspec} NPIPE, CMB lensing, and DESI BAO), which is not statistically significant. Moreover, by comparing the rightmost two columns of Table~\ref{table:chi2_2}, when using the EDE model with Pantheon+, adding a SH0ES prior increases the $\chi^2$ of the fit to \texttt{Camspec} NPIPE by $4.9$ as compared with a run that does not include the SH0ES prior. For a model to successfully resolve the Hubble tension, it must not worsen the fit to \emph{Planck} when imposing the SH0ES prior. 

When using pre-DESI BAO that prefers a slightly higher $\Omega_m$ instead of DESI BAO, the EDE bounds tighten again to $f_\mathrm{EDE} < 0.074 \: (95\% \,\, \mathrm{CL})$. One of the main qualitative conclusions of this \emph{Letter} is that the constraints on EDE are robust to the BAO dataset used, and  inclusion of BAO does not tighten the constraints on EDE parameters significantly compared to what CMB and CMB lensing already achieve.

Unlike \cite{PhysRevD.105.123536,Poulin:2021bjr,LaPosta:2021pgm, Smith:2022}, we do not find any hint for non-zero EDE using CMB and BAO data. In \cite{PhysRevD.105.123536}, the combination of ACT DR4 high-$\ell$ TT/TE/EE~\cite{ACT:2020frw,ACT:2020gnv} with \textit{Planck} 2018 low-$\ell$ and \textit{Planck} 2018 CMB lensing and pre-DESI BAO gave an $\approx3\sigma$ hint of EDE, with $f_\mathrm{EDE}=0.091^{+0.020}_{-0.036}$ (68\% CL) while our baseline constraints using \textit{Planck} CMB NPIPE + ACT DR6+\textit{Planck} PR4 CMB lensing and  DESI BAO give an upper bound of $f_\mathrm{EDE}<0.091$ (95\% CL). Whether or not the mild preference of ACT DR4 for a non-zero $f_\mathrm{EDE}$ is a subtle systematic artifact or a sign for new physics will likely be elucidated with the upcoming ACT DR6 and SPT-3G~\cite{SPT-3G:2014dbx} CMB power spectra measurements.

An important note is that MCMC chain convergence must be handled carefully, as a lack of true convergence can produce artificially tight bounds on parameters. In this work, we have imposed more stringent convergence criteria, requiring a Gelman-Rubin threshold of $R-1<0.01$ for convergence compared with the $R-1<0.05$ ~\cite{efstathiou2023improved} and $R-1<0.03$ \cite{PhysRevD.105.123536} used in previous works.

Furthermore, our analysis demonstrates the capability of accelerated inference with neural network emulators to efficiently explore parameter spaces and derive robust constraints within a reasonable timeframe. Without emulators, the work presented here could not have been carried out within only a few weeks from the release of the DESI BAO data.

Previous work has investigated the use of a profile likelihood to mitigate prior-volume effects that may bias Bayesian inference in the EDE context \cite{Herold:2021, efstathiou2023improved} (though such effects were found to be minimal in \cite{efstathiou2023improved}). We note that emulators may complicate the convergence of a profile likelihood due to small numerical noise in the emulator outputs. We leave the investigation into the use of emulators with a profile likelihood to future work using tools such as those described in \cite{Nygaard:2022}.

\section{acknowledgments}
We are indebted to Antony Lewis for providing the relevant DESI and Pantheon+ likelihood files in Cobaya and for helpful discussions. We are also grateful to Erminia Calabrese, Mathew Madhavacheril, Emmanuel Schaan, Alessio Spurio-Mancini, Arthur Kosowsky, Julien Lesgourgues, and  George  Efstathiou for discussions. We acknowledge the use \texttt{GetDist} \cite{Lewis:2019xzd} for analysing and plotting MCMC results. 
FJQ and BDS acknowledge support from the European Research Council (ERC) under the European Union’s Horizon 2020 research and innovation programme (Grant agreement No. 851274). BDS further acknowledges support from an STFC Ernest Rutherford Fellowship. KS acknowledges support from the  National Science Foundation Graduate Research Fellowship Program under Grant No. DGE 2036197. JCH acknowledges support from NSF grant AST-2108536, DOE grant DE-SC0011941, NASA grants 21-ATP21-0129 and 22-ADAP22-0145, the Sloan Foundation, and the Simons Foundation. Computations were performed on the [systemname] supercomputer at the SciNet HPC Consortium. SciNet is funded by Innovation, Science and Economic Development Canada; the Digital Research Alliance of Canada; the Ontario Research Fund: Research Excellence; and the University of Toronto.  We acknowledge the use of computational resources at the Flatiron Institute. We thank Nick Carriero, Robert Blackwell, and Dylan Simon from the Flatiron Institute for key computational advice.  The Flatiron Institute is supported by the Simons Foundation. We also acknowledge the use of computational resources from Columbia University's Shared Research Computing Facility project, which is supported by NIH Research Facility Improvement Grant 1G20RR030893-01, and associated funds from the New York State Empire State Development, Division of Science Technology and Innovation (NYSTAR) Contract C090171, both awarded April 15, 2010.  FJQ further acknowledges Cindy Cao for nice hospitality.

\appendix

\section{Priors used} \label{app.prior}

We sample the parameter space spanned by $\{f_\mathrm{EDE},\mathrm{log}_{10}(z_c),\theta_i,\ln (10^{10}A_s),H_0,n_s,\Omega_bh^2,\Omega_ch^2,\tau\}$. As pointed out in \cite{PhysRevD.105.123536,LaPosta:2021pgm}, the choice of prior range for $\log_{10}(z_c)$ is important because if it is extended to arbitrarily high redshifts, the parameter space is opened up, enabling $f_\mathrm{EDE}$ to take large values without having an impact on the CMB or other observables. Table \ref{table:priors} shows the priors used in this work. For the cases where we vary the neutrino mass sum, we adopt a broad uninformative uniform prior of $\sum{m_\nu}$ of [0,5] \, {\rm eV}.

\begin{table}[ht!]
\centering
\caption{Priors used in the EDE cosmological analysis of this work. Uniform priors are shown in square brackets and Gaussian priors with mean $\mu$ and standard deviation $\sigma$ are denoted $\mathcal{N}(\mu,\sigma)$. }
\begin{tabular}{cc}
\hline\hline
Parameter       & Prior      \\ \hline
$f_\mathrm{EDE}$ &   [0.001,0.5]   \\
$\mathrm{log}_{10}(z_c)$ &   [3,4.3]   \\
$\theta_i$ &   [0.1,3.1]   \\
$\ln (10^{10}A_s)$ &     [2.5,3.5]      \\ 
$H_0$           &   [50,90]        \\ 
$n_s$           &    [0.8,1.2] \\ 
$\Omega_bh^2$   & $\mathcal{N}(0.02233,0.00036)$\\ 
$\Omega_ch^2$   &  [0.005,0.99] \\ 
$\tau$          &     [0.01,0.8]     \\ \hline

\end{tabular}

\label{table:priors}
\end{table}

\section{Full marginalised posterior plots of different data-set combinations} \label{app:fullcontours}

\begin{figure*}
\includegraphics[width=\textwidth]{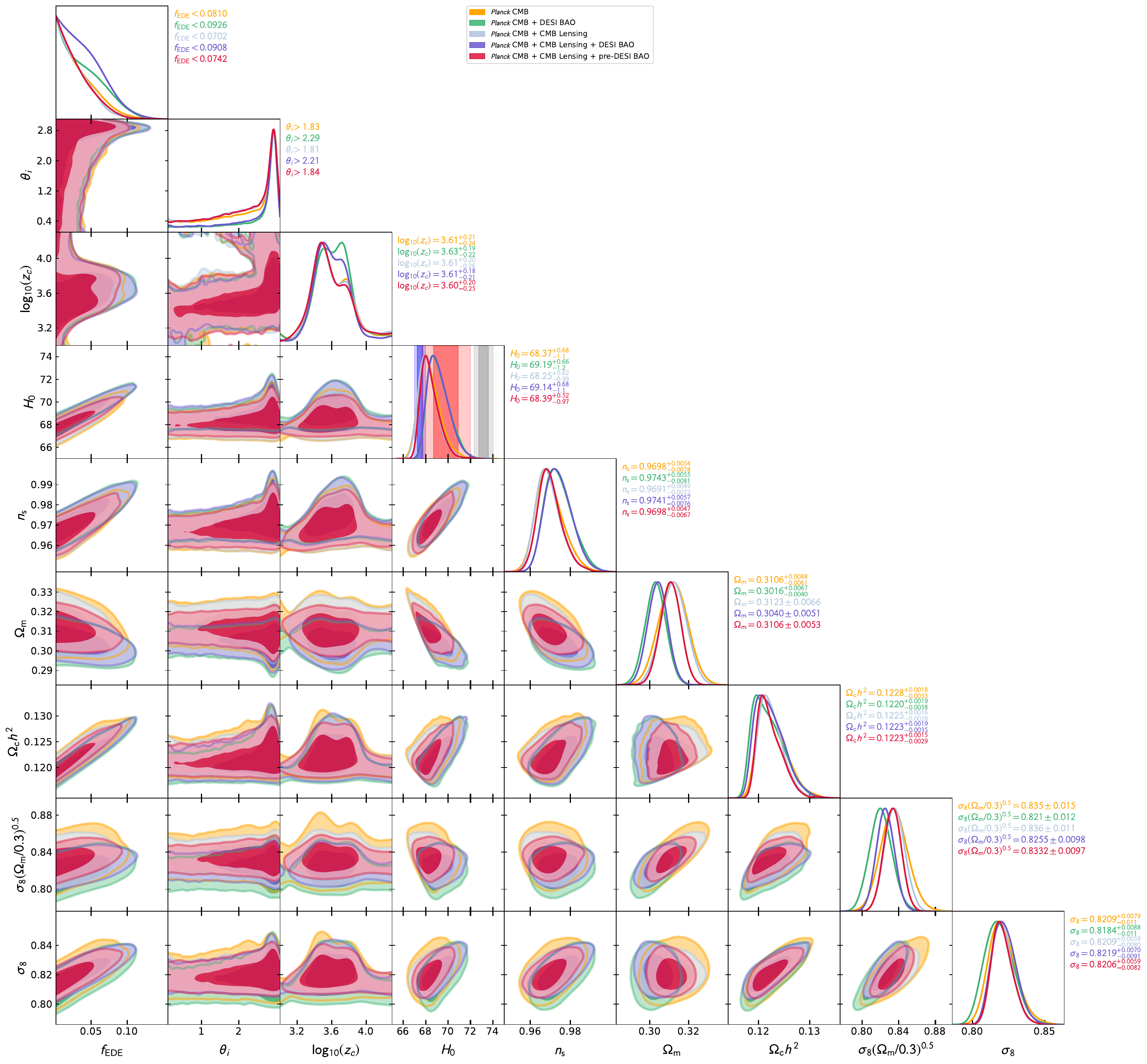}
\caption{Marginalized (1D and 2D) joint posterior probability distributions for the EDE parameters and a subset of other cosmological parameters of interest in fits to our baseline \textit{Planck} PR4 (NPIPE) TT+TE+EE data+ ACT DR6 and PR4 CMB lensing + DESI BAO data (blue). The vertical grey/magenta/blue bands in the $H_0$ panel show the latest SH0ES \cite{Scolnic_2022}, TRGB \cite{Freedman_2021} and \textit{Planck} constraints respectively. See Table~\ref{table:EDE-params-full} for a summary of these constraints and Subsection~\ref{res.1} for discussion.}
\label{fig:full_contours}
\end{figure*}

We show in Fig.~\ref{fig:full_contours} the full marginalized posteriors of the different subsets of the datasets used in the main analysis. For visualization purposes, we also show $95\% \; \mathrm{CL}$ bands centered at the mean value of the latest SH0ES, TRGB and \textit{Planck} measurements. None of the data combinations used are enough to bring the value of $H_0$ up to fully resolve the Hubble tension.

\section{Robustness tests} \label{app:b} 
\subsection{Emulators for EDE}

Our emulators are constructed with \texttt{cosmopower} \cite{SpurioMancini:2021ppk}, a wrapper for \texttt{TensorFlow} optimized for cosmological applications. The architecture of the neural networks and details on how they are produced can be found in \cite{SpurioMancini:2021ppk,Bolliet:2023sst}. The emulators for CMB spectra and distances were trained on 196091 samples spread in a Latin hypercube spanning the parameter space (with a test-train split of 80\%).  The input layer of the neural network emulators is the set of 6 $\Lambda$CDM parameters, namely, $A_s,n_s,\tau_\mathrm{reio},\Omega_c h^2, \Omega_b h^2$ and $H_0$ supplemented by the neutrino mass, the number of effective relativistic degrees of freedom in the early universe $N_\mathrm{eff}$, and the tensor-to-scalar ratio $r$. The generation of training data was done using a version of the Boltzmann code \texttt{class} \cite{2011arXiv1104.2932L,2011JCAP...07..034B} adapted to EDE models, \texttt{class\_ede} \footnote{\href{https://github.com/mwt5345/class_ede}{https://github.com/mwt5345/class\_ede}} \cite{Hill:2020osr}. We used the version of the code corresponding to the commit of Feb 16th 2023 on the GitHub repository\footnote{\href{https://github.com/mwt5345/class_ede/commit/199fbab08a5545c9f478c8137a1348c824d4874f}{199fbab08a5545c9f478c8137a1348c824d4874f}}, which is based on version v2.8.2 of \texttt{class}. We note that this \texttt{class} version could not allow extremely high-accuracy computation of the CMB high-$\ell$  regime because of its treatment of the Limber approximation for lensing. Hence, our current emulators will likely be obsolete in the Stage IV era. Nonetheless, as shown hereafter, the accuracy of our emulators is sufficient for the data analysis carried out in this paper. 
The precision settings adopted for the generation of the training data are as follows: 
\begin{itemize}
    \item \texttt{perturb\_sampling\_stepsize} : 0.05
    \item \texttt{neglect\_CMB\_sources\_below\_visibility}: 1e-30
    \item \texttt{transfer\_neglect\_late\_source}: 3000
    \item \texttt{halofit\_k\_per\_decade}: 3000
    \item \texttt{accurate\_lensing}: 1
    \item \texttt{num\_mu\_minus\_lmax} : 1000
    \item \texttt{delta\_l\_max}: 1000
    \item \texttt{k\_min\_tau0}: 0.002
    \item \texttt{k\_max\_tau0\_over\_l\_max}: 3.
      \item \texttt{k\_step\_sub}: 0.015
      \item \texttt{k\_step\_super}: 0.0001
      \item \texttt{k\_step\_super\_reduction}: 0.1
      \item \texttt{P\_k\_max\_h/Mpc}: 55/$h$
      \item \texttt{l\_max\_scalars}: 11000
\end{itemize}
These settings are the same as in  \cite{PhysRevD.105.123536} except that we do not set \texttt{l\_switch\_limber} = 40  but \texttt{perturb\_sampling\_stepsize} as indicated instead. The other difference is that we use $55/h$ rather than $100$ for \texttt{P\_k\_max\_h/Mpc}. These settings are motivated by the accuracy settings investigations carried out in \cite{McCarthy:2021lfp,PhysRevD.105.123536,Bolliet:2023sst}.

The CMB temperature and polarization spectra, and lensing potential power spectra, cover the multipole $2\leq \ell \leq 11,000$. Along with these, three redshift-dependent quantities are emulated over a redshift range $0<z<20$, namely, the Hubble parameter $H(z)$, the angular diameter distance $D_A(z)$, and the root-mean-square of the matter overdensity field smoothed over a spherical region of radius 8 Mpc, $\sigma_8(z)$. These redshift-dependent quantities constitute the building blocks for the theoretical prediction of BAO distance and RSD measurements. We also record and emulate a set of 16 derived parameters such as $\sigma_8$ (at $z=0$) or the primordial helium fraction. Our recombination and Big Bang Nucleosynthesis models correspond to the current fiducial settings of \texttt{class\_ede}:  \texttt{RECFAST} \citep{Seager:1999bc,2009MNRAS.397..445S,2012MNRAS.423.3227C} and \texttt{Parthenope} \citep{Consiglio:2017pot} with a fiducial $N_\mathrm{eff}=3.046$ using v1.2 of the code for a neutron lifetime of 880.2 s, identical to standard assumptions of \textit{Planck} 2017 papers, respectively. Nonlinear matter clustering is modeled with \texttt{hmcode} following implementation from \cite{Mead:2016zqy} and with CDM-only prescription, i.e., $\eta_0 = 0.603$ and $c_\mathrm{min}=3.13$ (following \texttt{class} notations).

\subsection{DESI BAO likelihood} \label{app.desi}
 
To test the likelihood, we reproduce benchmark constraints from \cite{desicollaboration2024desi}. We use the latest DESI BAO data \cite{desicollaboration2024desi} and its associated likelihood publicly available in the Markov Chain Monte Carlo (MCMC) sampler \texttt{cobaya} package \cite{cobaya}. This likelihood should correspond exactly to the one used in \cite{desicollaboration2024desi}. To confirm this, we reproduce the constraints in the first line of Table 4 of \cite{desicollaboration2024desi}, with DESI Y1 BAO data in combination with CMB. The CMB data is made of \texttt{planck\_2018\_lowl.TT}, \texttt{planck\_2018\_lowl.EE\_sroll2}, \texttt{planck\_NPIPE\_highl\_CamSpec.TTTEEE} as well as CMB lensing from ACT DR6 and NPIPE (without the inclusion of the normalization correction\footnote{DESI Y1 analysis used a version of the DR6 lensing likelihood release where this correction is effectively not applied, although the effect of applying versus not applying is very small for cosmological parameters of interest. }). The BAO data corresponds to the baseline, i.e., our \textbf{DESI BAO} (see main text). In \cite{desicollaboration2024desi} this data is analyzed within $\Lambda$CDM with three degenerate massive neutrinos and $N_\mathrm{eff}=3.044$, finding 
\begin{eqnarray}
    \Omega_\mathrm{m}=0.3037 \pm 0.0053 \, (68\% \, \mathrm{ CL}) \nonumber\\
    h = 0.6827\pm0.0042 \, (68\% \, \mathrm{ CL}) \nonumber\\ 
    \Sigma m_\nu<0.072   \, \mathrm{ eV} \, (95\% \, \mathrm{ CL}) \, \nonumber\\
\end{eqnarray}

First, we analyze this data with the $\Lambda$CDM emulator from \cite{Bolliet:2023sst} with three degenerate massive neutrinos and $N_\mathrm{eff}=3.046$. We find 
\begin{eqnarray}
    \Omega_\mathrm{m}=0.3037 \pm 0.0050 \, (68\% \, \mathrm{ CL}) \nonumber\\
    h = 0.6817 \pm 0.0040 \, (68\% \, \mathrm{ CL}) \nonumber\\ 
    \Sigma m_\nu<0.0735 \, \mathrm{ eV} \, (95\% \, \mathrm{ CL}) \, \nonumber\\
\end{eqnarray}
Thus, our result is $0.24 \sigma$ below the \cite{desicollaboration2024desi} result for $h$, and we obtain a $\approx 2 \%$ lower value for the 95\% CL upper limit on $\Sigma m_\nu$. We recover their value for $\Omega_\mathrm{m}$ to exact precision. This shows that the DESI likelihood we are using in this work is fully consistent with the one used in the official DESI paper. 

\begin{figure}
\centering
\includegraphics[width=\columnwidth]
{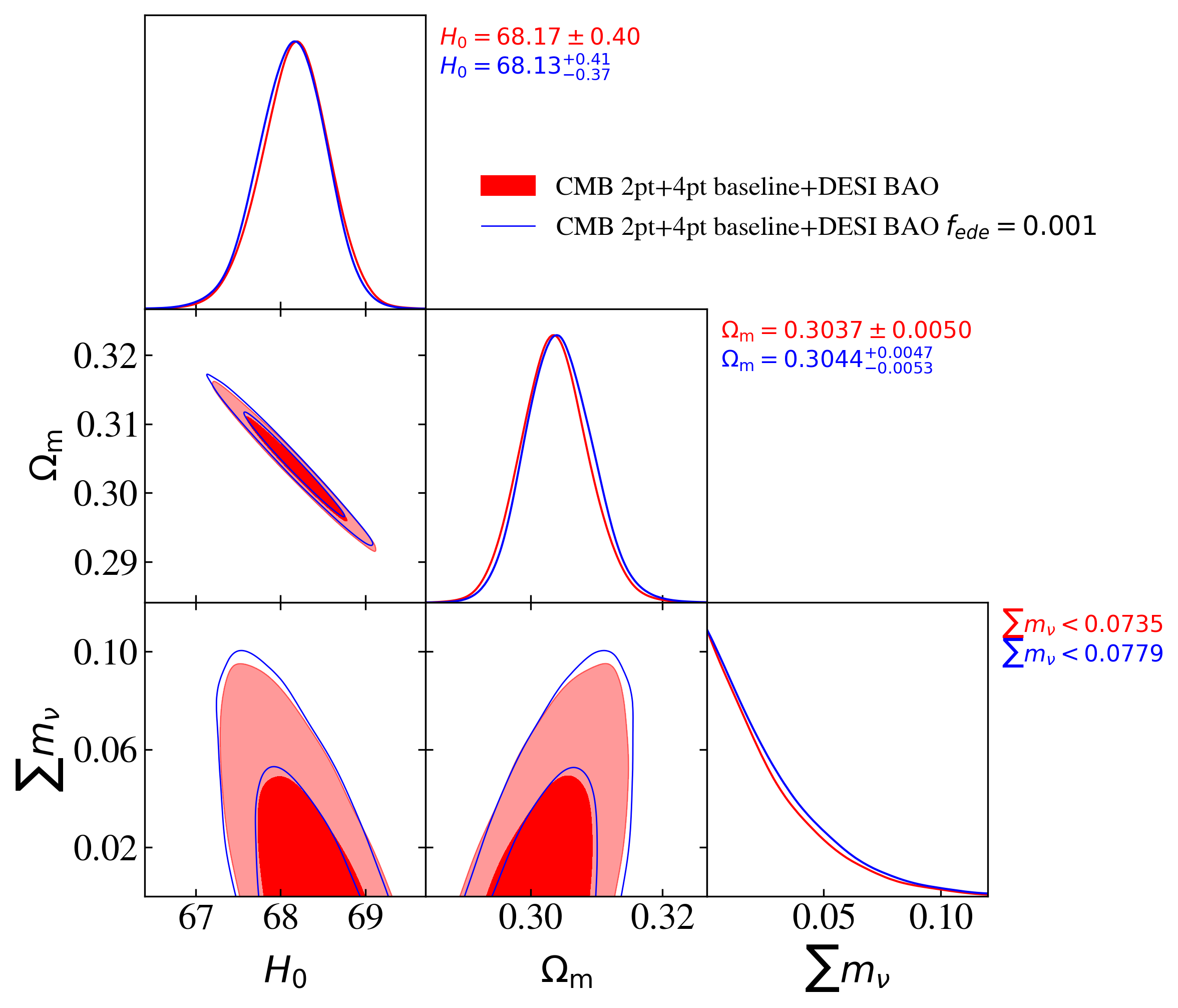}
 \caption{
Marginalized (1D and 2D) joint posterior probability distributions for $\sum{m_\nu}$, $H_0$ and $\Omega_m$ from the combination of CMB primary, CMB lensing and DESI BAO with $f_\mathrm{EDE}=0.001$. See text for details.}
\label{benchmark-desibao}
\end{figure}

In a second step, we analyze the same data combination with the EDE emulators, but setting the lowest possible amount of EDE, namely $f_\mathrm{EDE} = 0.001$ (\texttt{class\_ede} does not allow for a lower value of $f_\mathrm{EDE}$) and the other EDE parameters set to $\log_{10} z_c = 3.72$ and $\theta_{i,\mathrm{scf}}= 2.97$. These values for $z_c$ and $\theta_{i,\mathrm{scf}}$ are the best fit values from Table II of \cite{PhysRevD.105.123536}.
We find 
\begin{eqnarray}
    \Omega_\mathrm{m}=0.3044^{+0.0047}_{-0.0053} \, (68\% \, \mathrm{ CL}) \nonumber\\
    h = 0.6813^{+0.0041}_{-0.0037} \, (68\% \, \mathrm{ CL}) \nonumber\\ 
    \Sigma m_\nu<0.0779 \,  \mathrm{ eV} \,  (95\% \, \mathrm{ CL}) \,  \nonumber\\
\end{eqnarray}
Thus, these EDE emulator results are $0.14 \sigma$ above and $0.1 \sigma$ below the non-EDE emulator results (from the previous paragraph) for $\Omega_\mathrm{m}$ and $h$, respectively. The 95\% CL neutrino mass sum limit also increases by $\approx 6\%$. These differences are likely explained by the fact that there is still a small amount of EDE. But again the differences are a small fraction of the uncertainties, which shows that our EDE emulators are suited for analyzing current DESI BAO data without any statistically significant bias. 

Contours for both analyses described here are shown in Fig.~\ref{benchmark-desibao}.

\subsection{ACTDR4 TTTEEE EDE constraints benchmark}

To validate our emulators at the level of CMB temperature and polarization spectra, we reproduce EDE constraints from \cite{PhysRevD.105.123536}. We use the public, foreground marginalized, likelihood code \texttt{pyactlike}\footnote{\href{https://github.com/ACTCollaboration/pyactlike}{https://github.com/ACTCollaboration/pyactlike}}. The data are stored in the same online repository and presented in detail in \cite{ACT:2020frw,ACT:2020gnv}.  We reproduce results corresponding to the last two columns of Table II of \cite{PhysRevD.105.123536}, which use ACT DR4 TT+TE+EE spectra along with a Gaussian prior on the optical depth $\tau = 0.065\pm 0.015$ (mean and standard deviation). Although these results were obtained with the same EDE code, \texttt{class\_ede}, there are minor differences between our emulator settings and the settings of \cite{PhysRevD.105.123536}. In addition to slightly different precision settings (see above), for the nonlinear evolution we use \texttt{hmcode} while use \cite{PhysRevD.105.123536} \texttt{halofit}, and for neutrinos, we use three degenerate  massive neutrinos, while \cite{PhysRevD.105.123536} used one massive and two massless neutrinos. Furthermore,  \cite{PhysRevD.105.123536} required a convergence criterion of $R-1<0.03$ while we require at least $R-1<0.01$. By re-analyzing the chains from \cite{PhysRevD.105.123536} (publicly available online\footnote{\href{https://users.flatironinstitute.org/~chill/H21_data/}{https://flatironinstitute.org/chill/H21\_data/}}), excluding 10\% of burn-in, we get a convergence criterion of $R-1=0.0117$. In comparison, our chains have $R-1=0.0023$ (excluding 20\% of burn-in). 

\begin{figure*}
\includegraphics[width=\textwidth]{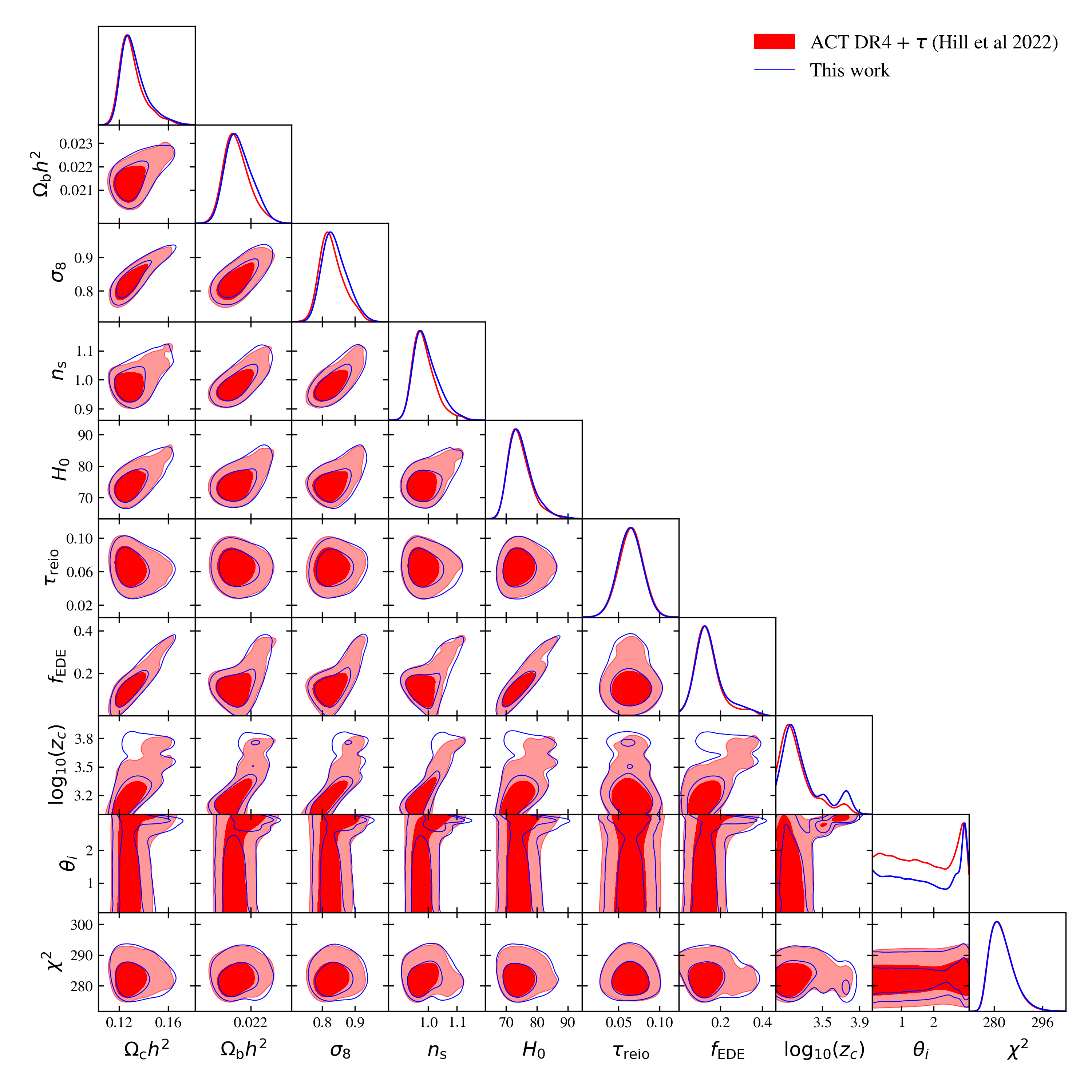}
\caption{Marginalized (1D and 2D) joint posterior probability distributions of  EDE parameters and a subset of other parameters in fits of ACT DR4 CMB data in combination with a Gaussian prior on the optical depth. See text for details.}
\label{fig:pyactlike}
\end{figure*}

As shown in Fig.~\ref{fig:pyactlike}, the marginalized joint posterior probability distributions are almost identical between the analysis from \cite{PhysRevD.105.123536} and our recovery run. Slight differences in the tails can be attributed to emulator accuracy as well as the different settings between both analyses mentioned above. 

\begin{figure}
\centering
\includegraphics[width=\columnwidth]
{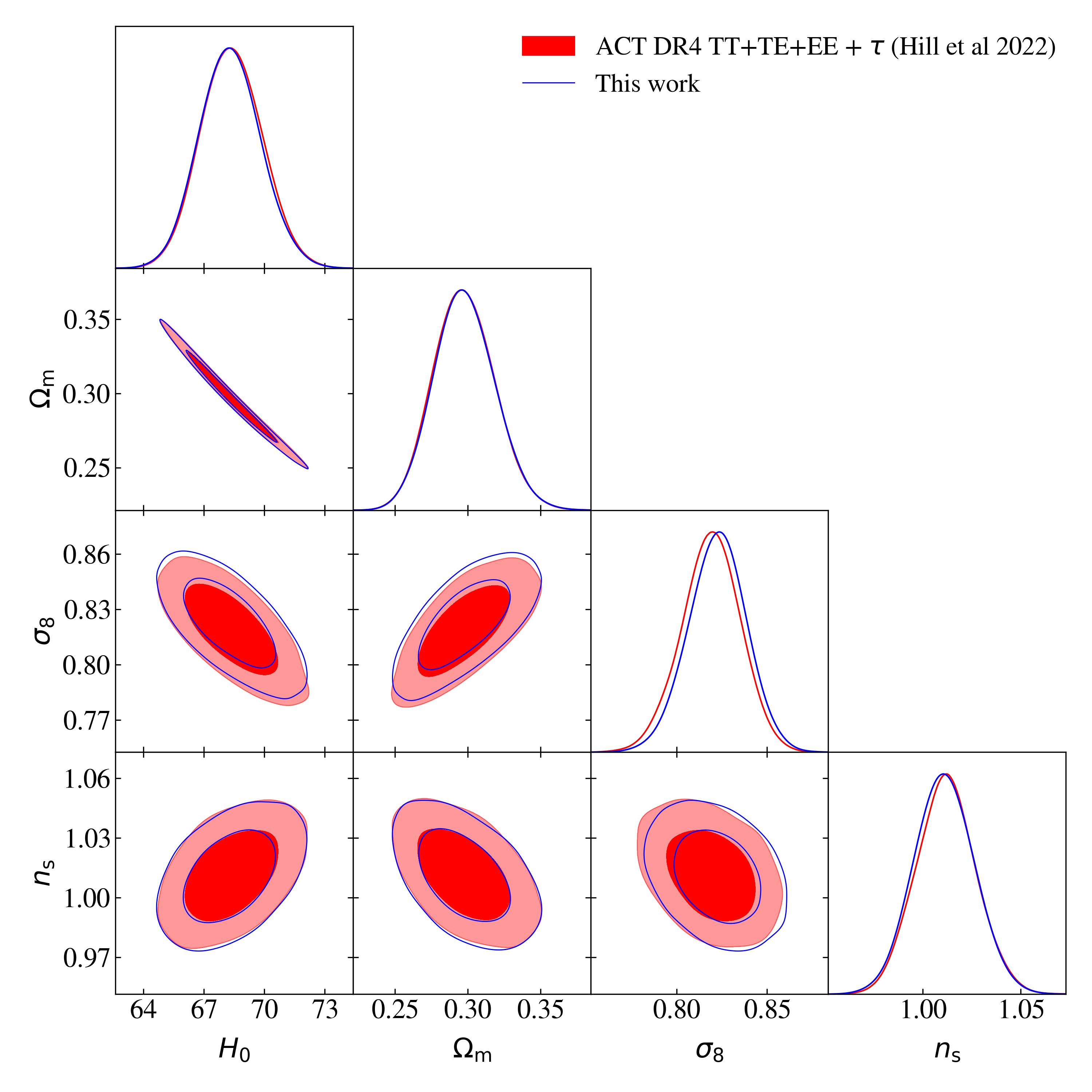}
\caption{Marginalized (1D and 2D) joint posterior probability distributions of $\Lambda$CDM parameters in fits of ACT DR4 CMB data in combination with a Gaussian prior on the optical depth. The blue contours use EDE emulators with minimal EDE content, $f_\mathrm{EDE}=0.001$.  See text for details. The slight shift in $\sigma_8$ is in the expected direction, and the positive degeneracy between $f_{\mathrm{EDE}}$ and $\sigma_8$ could also be due to slightly different accuracy settings.}
\label{fig:pyactlikelcdm}
\end{figure}

We also perform a test of $\Lambda$CDM constraints using an EDE emulator with a minimal amount of EDE. Contours are shown in Fig.~\ref{fig:pyactlikelcdm}.

Overall, these results validate the use of the EDE emulators for the analysis of ACT DR4 CMB TT+TE+EE spectra.

\begin{figure}
\centering
\includegraphics[width=\columnwidth]
{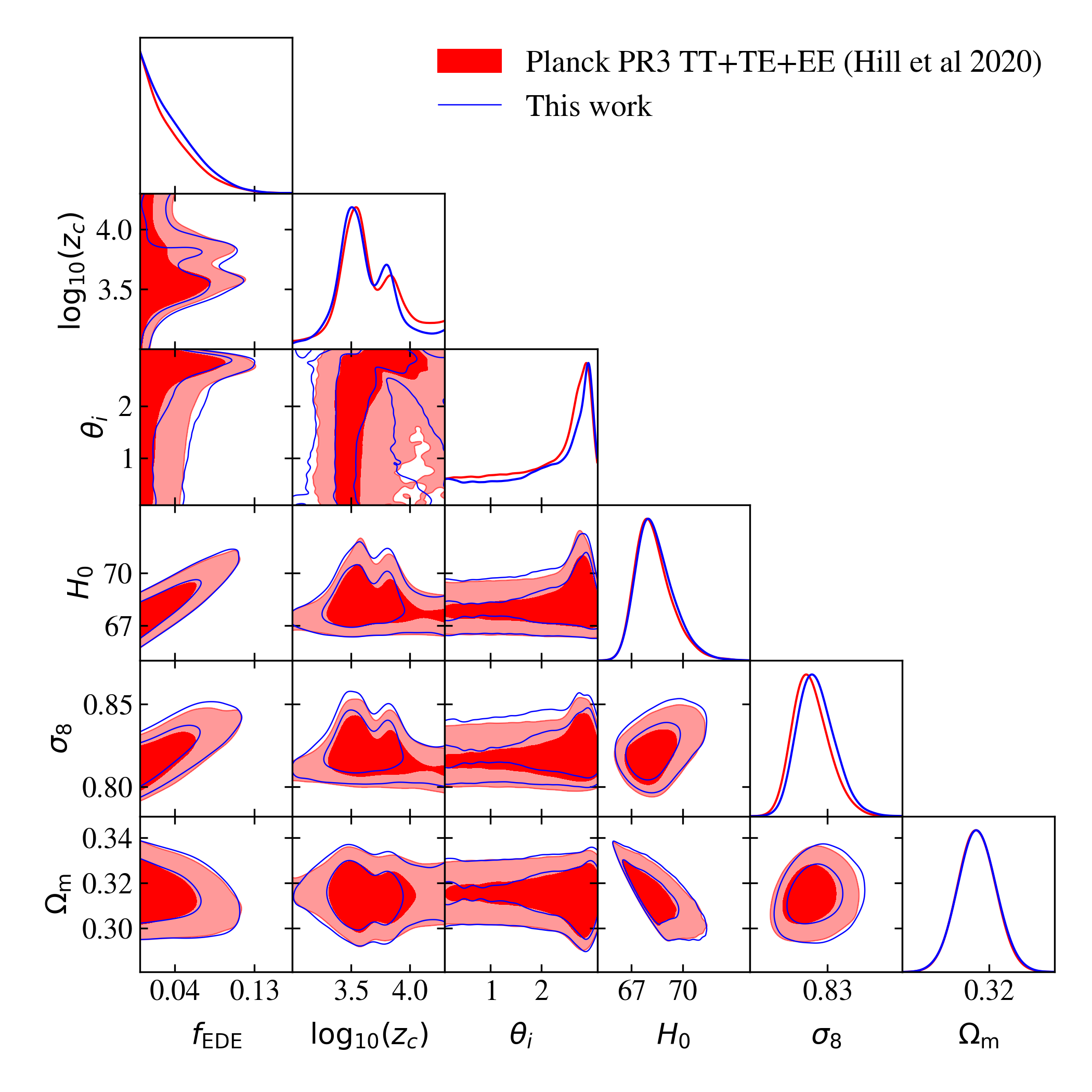}
 \caption{
Marginalized (1D and 2D) joint posterior probability distributions for \textit{Planck} PR3 benchmark. See text for details.}
\label{benchmark-pr3}
\end{figure}

\subsection{Planck 2018 TTTEE EDE constraints}
\label{app.planck19benchmark}

For completeness, we reproduce results from \cite{Hill:2020osr} corresponding to the first column of their Table I. For this analysis, \cite{Hill:2020osr} uses CMB data from \textit{Planck} PR3 including \texttt{planck\_2018\_lowl.TT}, \texttt{planck\_2018\_lowl.EE} and \texttt{planck\_2018\_highl\_plik.TTTEEE} as well as a Gaussian prior on the Sunyaev-Zel'dovich components. While \cite{Hill:2020osr} carries out the analysis using the \texttt{plik} high-$\ell$ likelihood, we choose to use the native Python implementation available in \texttt{cobaya}, i.e., \texttt{planck\_2018\_highl\_plik.TTTEEE\_lite\_native}\footnote{See \href{https://cobaya.readthedocs.io/en/latest/likelihood_planck.html}{https://cobaya.readthedocs.io/likelihood\_planck.html}.}.

\begin{figure}
\centering
\includegraphics[width=\columnwidth]
{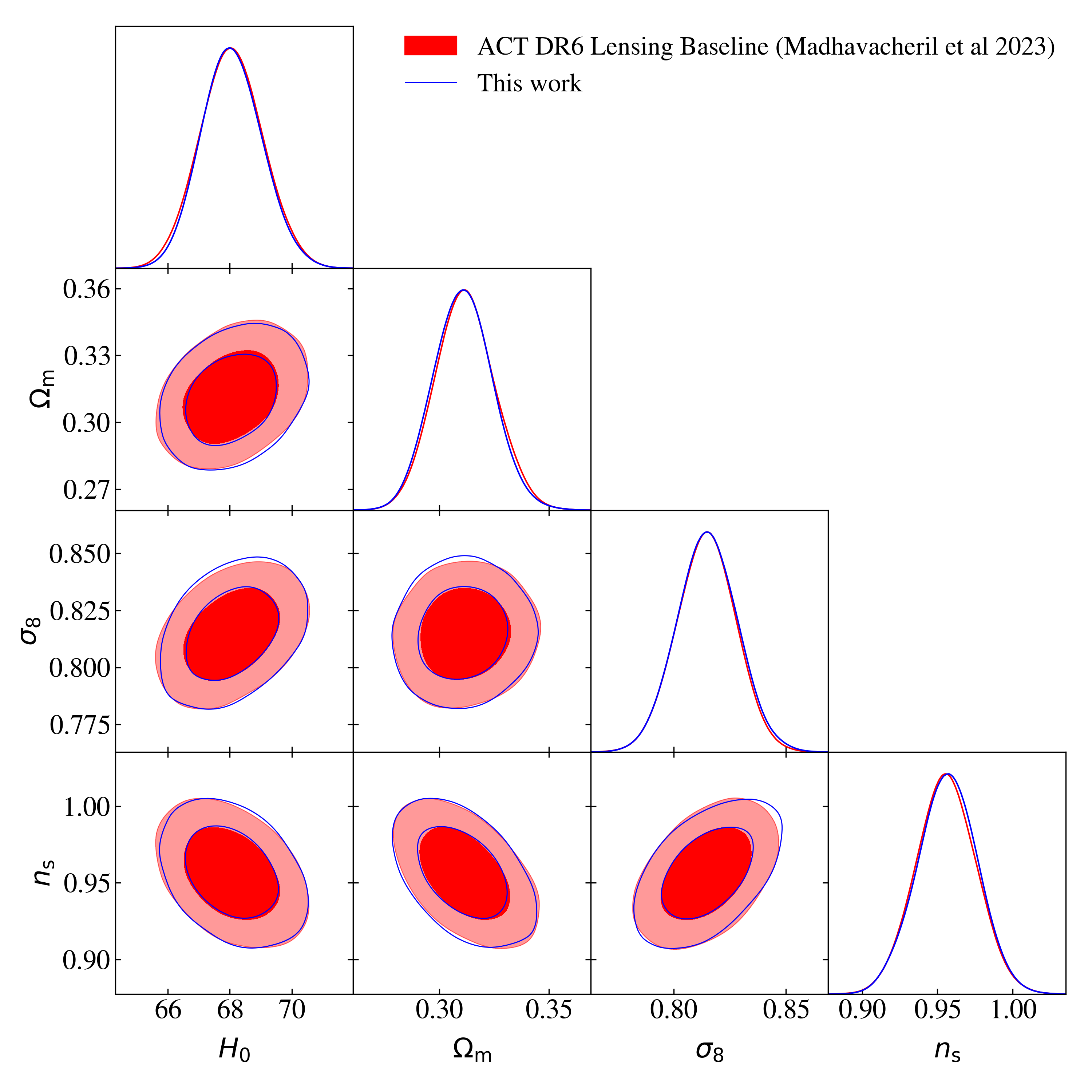}
 \caption{
Marginalized (1D and 2D) joint posterior probability distributions for ACT DR6 lensing benchmark. This includes pre-DESI BAO data. The blue contours use EDE emulators with minimal EDE content, $f_\mathrm{EDE}=0.001$. See text for details.}
\label{benchmark-dr6lens}
\end{figure}

The authors of  \cite{Hill:2020osr} require a convergence criterion of $R-1<0.05$ and use an earlier version of \texttt{class\_ede} than the one on which our EDE emulators are based. They consider one massive and two massless neutrinos and default \texttt{class} v2.8.2 settings for other cosmological and precision parameters, except for \texttt{P\_k\_max\_h/Mpc} which they set to 20. Thus, our accuracy settings and convergence criterion are considerably more demanding than those of \cite{Hill:2020osr}.

The chains from \cite{Hill:2020osr} are available online\footnote{\href{https://users.flatironinstitute.org/~chill/H20_data/}{flatironinstitute.org/chill/H20\_data/}}. Analysing these and excluding 50\% of burn-in we get $R-1=0.02605$. In comparison, excluding the same burn-in fraction, our chains have $R-1=0.00099$. Corresponding contours are shown on Fig.~\ref{benchmark-pr3}. In spite of the differences mentioned above, both analyses match nearly perfectly. We get $f_\mathrm{EDE} < 0.0925$ (95\% CL) (against 0.0908 from our re-analysis of the chains from \cite{Hill:2020osr}; we note that this is slightly different from the value quoted in their table, 0.087), $\mathrm{log}_{10}(z_c) = 3.63\pm 0.24$ (68\% CL) (against $\mathrm{log}_{10}(z_c) = 3.66^{+0.24}_{-0.28}$ from our re-analysis of the chains from \cite{Hill:2020osr}), $\theta_i > 1.73$  (68\% CL) (against 1.67 from our re-analysis of the chains from \cite{Hill:2020osr}), $\Omega_\mathrm{m} = 0.3146\pm 0.0088$ (68\%CL) (against $0.3144\pm 0.0086$  from our re-analysis of the chains from \cite{Hill:2020osr}), $\sigma_8 = 0.8235^{+0.0096}_{-0.013}$ (68\%CL) (against $0.8202^{+0.0091}_{-0.013}$ from our re-analysis of the chains from \cite{Hill:2020osr}) and $H_0 = 68.42^{+0.79}_{-1.3} \; \Hunit$ (68\% CL) (against $68.33^{+0.73}_{-1.3}$ from our re-analysis of the chains from \cite{Hill:2020osr}).

\subsection{ACT DR6 lensing with EDE emulator $\Lambda$CDM constraints benchmark}
To test our EDE emulators on CMB lensing data, we perform a baseline ACT DR6 lensing analysis with EDE emulators and minimal amount of EDE. In particular, we set $f_\mathrm{EDE}=0.001$, $\theta_{i,\mathrm{scf}}=2.97$ and $\log_{10} z_c = 3.72$. The data include DR6 and \textit{Planck} lensing as well as pre-DESI BAO. Excluding 10\% of burn-in, our chains have $R-1=0.00586$. Contours are shown in Fig.~\ref{benchmark-dr6lens}.

\bibliography{apssamp}

\providecommand{\noopsort}[1]{}\providecommand{\singleletter}[1]{#1}%
\begin{thebibliography}{56}%
\makeatletter
\providecommand \@ifxundefined [1]{%
 \@ifx{#1\undefined}
}%
\providecommand \@ifnum [1]{%
 \ifnum #1\expandafter \@firstoftwo
 \else \expandafter \@secondoftwo
 \fi
}%
\providecommand \@ifx [1]{%
 \ifx #1\expandafter \@firstoftwo
 \else \expandafter \@secondoftwo
 \fi
}%
\providecommand \natexlab [1]{#1}%
\providecommand \enquote  [1]{``#1''}%
\providecommand \bibnamefont  [1]{#1}%
\providecommand \bibfnamefont [1]{#1}%
\providecommand \citenamefont [1]{#1}%
\providecommand \href@noop [0]{\@secondoftwo}%
\providecommand \href [0]{\begingroup \@sanitize@url \@href}%
\providecommand \@href[1]{\@@startlink{#1}\@@href}%
\providecommand \@@href[1]{\endgroup#1\@@endlink}%
\providecommand \@sanitize@url [0]{\catcode `\\12\catcode `\$12\catcode `\&12\catcode `\#12\catcode `\^12\catcode `\_12\catcode `\%12\relax}%
\providecommand \@@startlink[1]{}%
\providecommand \@@endlink[0]{}%
\providecommand \url  [0]{\begingroup\@sanitize@url \@url }%
\providecommand \@url [1]{\endgroup\@href {#1}{\urlprefix }}%
\providecommand \urlprefix  [0]{URL }%
\providecommand \Eprint [0]{\href }%
\providecommand \doibase [0]{http://dx.doi.org/}%
\providecommand \selectlanguage [0]{\@gobble}%
\providecommand \bibinfo  [0]{\@secondoftwo}%
\providecommand \bibfield  [0]{\@secondoftwo}%
\providecommand \translation [1]{[#1]}%
\providecommand \BibitemOpen [0]{}%
\providecommand \bibitemStop [0]{}%
\providecommand \bibitemNoStop [0]{.\EOS\space}%
\providecommand \EOS [0]{\spacefactor3000\relax}%
\providecommand \BibitemShut  [1]{\csname bibitem#1\endcsname}%
\let\auto@bib@innerbib\@empty
\bibitem [{\citenamefont {Aghanim}\ \emph {et~al.}(2020)\citenamefont {Aghanim} \emph {et~al.}}]{Planck:2018_cosmo_params}%
  \BibitemOpen
  \bibfield  {author} {\bibinfo {author} {\bibfnamefont {N.}~\bibnamefont {Aghanim}} \emph {et~al.} (\bibinfo {collaboration} {Planck}),\ }\href {\doibase 10.1051/0004-6361/201833910} {\bibfield  {journal} {\bibinfo  {journal} {Astron. Astrophys.}\ }\textbf {\bibinfo {volume} {641}},\ \bibinfo {pages} {A6} (\bibinfo {year} {2020})},\ \bibinfo {note} {[Erratum: Astron.Astrophys. 652, C4 (2021)]},\ \Eprint {http://arxiv.org/abs/1807.06209} {arXiv:1807.06209 [astro-ph.CO]} \BibitemShut {NoStop}%
\bibitem [{\citenamefont {Breuval}\ \emph {et~al.}(2024)\citenamefont {Breuval}, \citenamefont {Riess}, \citenamefont {Casertano}, \citenamefont {Yuan}, \citenamefont {Macri}, \citenamefont {Romaniello}, \citenamefont {Murakami}, \citenamefont {Scolnic}, \citenamefont {Anand},\ and\ \citenamefont {Soszyński}}]{Breuval:2024lsv}%
  \BibitemOpen
  \bibfield  {author} {\bibinfo {author} {\bibfnamefont {L.}~\bibnamefont {Breuval}}, \bibinfo {author} {\bibfnamefont {A.~G.}\ \bibnamefont {Riess}}, \bibinfo {author} {\bibfnamefont {S.}~\bibnamefont {Casertano}}, \bibinfo {author} {\bibfnamefont {W.}~\bibnamefont {Yuan}}, \bibinfo {author} {\bibfnamefont {L.~M.}\ \bibnamefont {Macri}}, \bibinfo {author} {\bibfnamefont {M.}~\bibnamefont {Romaniello}}, \bibinfo {author} {\bibfnamefont {Y.~S.}\ \bibnamefont {Murakami}}, \bibinfo {author} {\bibfnamefont {D.}~\bibnamefont {Scolnic}}, \bibinfo {author} {\bibfnamefont {G.~S.}\ \bibnamefont {Anand}}, \ and\ \bibinfo {author} {\bibfnamefont {I.}~\bibnamefont {Soszyński}},\ }\href@noop {} {\enquote {\bibinfo {title} {Small magellanic cloud cepheids observed with the hubble space telescope provide a new anchor for the sh0es distance ladder},}\ } (\bibinfo {year} {2024}),\ \Eprint {http://arxiv.org/abs/2404.08038} {arXiv:2404.08038 [astro-ph.CO]} \BibitemShut {NoStop}%
\bibitem [{\citenamefont {Verde}\ \emph {et~al.}(2019)\citenamefont {Verde}, \citenamefont {Treu},\ and\ \citenamefont {Riess}}]{Verde:2019}%
  \BibitemOpen
  \bibfield  {author} {\bibinfo {author} {\bibfnamefont {L.}~\bibnamefont {Verde}}, \bibinfo {author} {\bibfnamefont {T.}~\bibnamefont {Treu}}, \ and\ \bibinfo {author} {\bibfnamefont {A.~G.}\ \bibnamefont {Riess}},\ }\href {\doibase 10.1038/s41550-019-0902-0} {\bibfield  {journal} {\bibinfo  {journal} {Nature Astron.}\ }\textbf {\bibinfo {volume} {3}},\ \bibinfo {pages} {891} (\bibinfo {year} {2019})},\ \Eprint {http://arxiv.org/abs/1907.10625} {arXiv:1907.10625 [astro-ph.CO]} \BibitemShut {NoStop}%
\bibitem [{\citenamefont {Di~Valentino}\ \emph {et~al.}(2021)\citenamefont {Di~Valentino}, \citenamefont {Mena}, \citenamefont {Pan}, \citenamefont {Visinelli}, \citenamefont {Yang}, \citenamefont {Melchiorri}, \citenamefont {Mota}, \citenamefont {Riess},\ and\ \citenamefont {Silk}}]{DiValentino:2021}%
  \BibitemOpen
  \bibfield  {author} {\bibinfo {author} {\bibfnamefont {E.}~\bibnamefont {Di~Valentino}}, \bibinfo {author} {\bibfnamefont {O.}~\bibnamefont {Mena}}, \bibinfo {author} {\bibfnamefont {S.}~\bibnamefont {Pan}}, \bibinfo {author} {\bibfnamefont {L.}~\bibnamefont {Visinelli}}, \bibinfo {author} {\bibfnamefont {W.}~\bibnamefont {Yang}}, \bibinfo {author} {\bibfnamefont {A.}~\bibnamefont {Melchiorri}}, \bibinfo {author} {\bibfnamefont {D.~F.}\ \bibnamefont {Mota}}, \bibinfo {author} {\bibfnamefont {A.~G.}\ \bibnamefont {Riess}}, \ and\ \bibinfo {author} {\bibfnamefont {J.}~\bibnamefont {Silk}},\ }\href {\doibase 10.1088/1361-6382/ac086d} {\bibfield  {journal} {\bibinfo  {journal} {Class. Quant. Grav.}\ }\textbf {\bibinfo {volume} {38}},\ \bibinfo {pages} {153001} (\bibinfo {year} {2021})},\ \Eprint {http://arxiv.org/abs/2103.01183} {arXiv:2103.01183 [astro-ph.CO]} \BibitemShut {NoStop}%
\bibitem [{\citenamefont {Freedman}(2021)}]{Freedman_2021}%
  \BibitemOpen
  \bibfield  {author} {\bibinfo {author} {\bibfnamefont {W.~L.}\ \bibnamefont {Freedman}},\ }\href {\doibase 10.3847/1538-4357/ac0e95} {\bibfield  {journal} {\bibinfo  {journal} {The Astrophysical Journal}\ }\textbf {\bibinfo {volume} {919}},\ \bibinfo {pages} {16} (\bibinfo {year} {2021})}\BibitemShut {NoStop}%
\bibitem [{\citenamefont {{Freedman}}\ \emph {et~al.}(2001)\citenamefont {{Freedman}}, \citenamefont {{Madore}}, \citenamefont {{Gibson}}, \citenamefont {{Ferrarese}}, \citenamefont {{Kelson}}, \citenamefont {{Sakai}}, \citenamefont {{Mould}}, \citenamefont {{Kennicutt}}, \citenamefont {{Ford}}, \citenamefont {{Graham}}, \citenamefont {{Huchra}}, \citenamefont {{Hughes}}, \citenamefont {{Illingworth}}, \citenamefont {{Macri}},\ and\ \citenamefont {{Stetson}}}]{2001ApJ...553...47F}%
  \BibitemOpen
  \bibfield  {author} {\bibinfo {author} {\bibfnamefont {W.~L.}\ \bibnamefont {{Freedman}}}, \bibinfo {author} {\bibfnamefont {B.~F.}\ \bibnamefont {{Madore}}}, \bibinfo {author} {\bibfnamefont {B.~K.}\ \bibnamefont {{Gibson}}}, \bibinfo {author} {\bibfnamefont {L.}~\bibnamefont {{Ferrarese}}}, \bibinfo {author} {\bibfnamefont {D.~D.}\ \bibnamefont {{Kelson}}}, \bibinfo {author} {\bibfnamefont {S.}~\bibnamefont {{Sakai}}}, \bibinfo {author} {\bibfnamefont {J.~R.}\ \bibnamefont {{Mould}}}, \bibinfo {author} {\bibfnamefont {J.}~\bibnamefont {{Kennicutt}}, \bibfnamefont {Robert~C.}}, \bibinfo {author} {\bibfnamefont {H.~C.}\ \bibnamefont {{Ford}}}, \bibinfo {author} {\bibfnamefont {J.~A.}\ \bibnamefont {{Graham}}}, \bibinfo {author} {\bibfnamefont {J.~P.}\ \bibnamefont {{Huchra}}}, \bibinfo {author} {\bibfnamefont {S.~M.~G.}\ \bibnamefont {{Hughes}}}, \bibinfo {author} {\bibfnamefont {G.~D.}\ \bibnamefont {{Illingworth}}}, \bibinfo {author} {\bibfnamefont {L.~M.}\ \bibnamefont {{Macri}}}, \ and\ \bibinfo
  {author} {\bibfnamefont {P.~B.}\ \bibnamefont {{Stetson}}},\ }\href {\doibase 10.1086/320638} {\bibfield  {journal} {\bibinfo  {journal} {\apj}\ }\textbf {\bibinfo {volume} {553}},\ \bibinfo {pages} {47} (\bibinfo {year} {2001})},\ \Eprint {http://arxiv.org/abs/astro-ph/0012376} {arXiv:astro-ph/0012376 [astro-ph]} \BibitemShut {NoStop}%
\bibitem [{\citenamefont {Freedman}\ and\ \citenamefont {Madore}(2023)}]{Freedman:2023}%
  \BibitemOpen
  \bibfield  {author} {\bibinfo {author} {\bibfnamefont {W.~L.}\ \bibnamefont {Freedman}}\ and\ \bibinfo {author} {\bibfnamefont {B.~F.}\ \bibnamefont {Madore}},\ }\href {\doibase 10.1088/1475-7516/2023/11/050} {\bibfield  {journal} {\bibinfo  {journal} {JCAP}\ }\textbf {\bibinfo {volume} {11}},\ \bibinfo {pages} {050} (\bibinfo {year} {2023})},\ \Eprint {http://arxiv.org/abs/2309.05618} {arXiv:2309.05618 [astro-ph.CO]} \BibitemShut {NoStop}%
\bibitem [{\citenamefont {McDonough}\ \emph {et~al.}(2023)\citenamefont {McDonough}, \citenamefont {Hill}, \citenamefont {Ivanov}, \citenamefont {La~Posta},\ and\ \citenamefont {Toomey}}]{McDonough:2023qcu}%
  \BibitemOpen
  \bibfield  {author} {\bibinfo {author} {\bibfnamefont {E.}~\bibnamefont {McDonough}}, \bibinfo {author} {\bibfnamefont {J.~C.}\ \bibnamefont {Hill}}, \bibinfo {author} {\bibfnamefont {M.~M.}\ \bibnamefont {Ivanov}}, \bibinfo {author} {\bibfnamefont {A.}~\bibnamefont {La~Posta}}, \ and\ \bibinfo {author} {\bibfnamefont {M.~W.}\ \bibnamefont {Toomey}},\ }\href@noop {} {\bibfield  {journal} {\bibinfo  {journal} {International Journal of Modern Physics D}\ } (\bibinfo {year} {2023})},\ \Eprint {http://arxiv.org/abs/2310.19899} {arXiv:2310.19899 [astro-ph.CO]} \BibitemShut {NoStop}%
\bibitem [{\citenamefont {Poulin}\ \emph {et~al.}(2023)\citenamefont {Poulin}, \citenamefont {Smith},\ and\ \citenamefont {Karwal}}]{Poulin:2023lkg}%
  \BibitemOpen
  \bibfield  {author} {\bibinfo {author} {\bibfnamefont {V.}~\bibnamefont {Poulin}}, \bibinfo {author} {\bibfnamefont {T.~L.}\ \bibnamefont {Smith}}, \ and\ \bibinfo {author} {\bibfnamefont {T.}~\bibnamefont {Karwal}},\ }\href {\doibase 10.1016/j.dark.2023.101348} {\bibfield  {journal} {\bibinfo  {journal} {Phys. Dark Univ.}\ }\textbf {\bibinfo {volume} {42}},\ \bibinfo {pages} {101348} (\bibinfo {year} {2023})},\ \Eprint {http://arxiv.org/abs/2302.09032} {arXiv:2302.09032 [astro-ph.CO]} \BibitemShut {NoStop}%
\bibitem [{\citenamefont {Poulin}\ \emph {et~al.}(2019)\citenamefont {Poulin}, \citenamefont {Smith}, \citenamefont {Karwal},\ and\ \citenamefont {Kamionkowski}}]{Poulin2019}%
  \BibitemOpen
  \bibfield  {author} {\bibinfo {author} {\bibfnamefont {V.}~\bibnamefont {Poulin}}, \bibinfo {author} {\bibfnamefont {T.~L.}\ \bibnamefont {Smith}}, \bibinfo {author} {\bibfnamefont {T.}~\bibnamefont {Karwal}}, \ and\ \bibinfo {author} {\bibfnamefont {M.}~\bibnamefont {Kamionkowski}},\ }\href {\doibase 10.1103/PhysRevLett.122.221301} {\bibfield  {journal} {\bibinfo  {journal} {Phys. Rev. Lett.}\ }\textbf {\bibinfo {volume} {122}},\ \bibinfo {pages} {221301} (\bibinfo {year} {2019})}\BibitemShut {NoStop}%
\bibitem [{\citenamefont {Lin}\ \emph {et~al.}(2019)\citenamefont {Lin}, \citenamefont {Benevento}, \citenamefont {Hu},\ and\ \citenamefont {Raveri}}]{Lin:2019qug}%
  \BibitemOpen
  \bibfield  {author} {\bibinfo {author} {\bibfnamefont {M.-X.}\ \bibnamefont {Lin}}, \bibinfo {author} {\bibfnamefont {G.}~\bibnamefont {Benevento}}, \bibinfo {author} {\bibfnamefont {W.}~\bibnamefont {Hu}}, \ and\ \bibinfo {author} {\bibfnamefont {M.}~\bibnamefont {Raveri}},\ }\href {\doibase 10.1103/PhysRevD.100.063542} {\bibfield  {journal} {\bibinfo  {journal} {Phys. Rev. D}\ }\textbf {\bibinfo {volume} {100}},\ \bibinfo {pages} {063542} (\bibinfo {year} {2019})},\ \Eprint {http://arxiv.org/abs/1905.12618} {arXiv:1905.12618 [astro-ph.CO]} \BibitemShut {NoStop}%
\bibitem [{\citenamefont {Agrawal}\ \emph {et~al.}(2023)\citenamefont {Agrawal}, \citenamefont {Cyr-Racine}, \citenamefont {Pinner},\ and\ \citenamefont {Randall}}]{Agrawal:2019lmo}%
  \BibitemOpen
  \bibfield  {author} {\bibinfo {author} {\bibfnamefont {P.}~\bibnamefont {Agrawal}}, \bibinfo {author} {\bibfnamefont {F.-Y.}\ \bibnamefont {Cyr-Racine}}, \bibinfo {author} {\bibfnamefont {D.}~\bibnamefont {Pinner}}, \ and\ \bibinfo {author} {\bibfnamefont {L.}~\bibnamefont {Randall}},\ }\href {\doibase 10.1016/j.dark.2023.101347} {\bibfield  {journal} {\bibinfo  {journal} {Phys. Dark Univ.}\ }\textbf {\bibinfo {volume} {42}},\ \bibinfo {pages} {101347} (\bibinfo {year} {2023})},\ \Eprint {http://arxiv.org/abs/1904.01016} {arXiv:1904.01016 [astro-ph.CO]} \BibitemShut {NoStop}%
\bibitem [{\citenamefont {Kamionkowski}\ and\ \citenamefont {Riess}(2023)}]{Kamionkowski:2022}%
  \BibitemOpen
  \bibfield  {author} {\bibinfo {author} {\bibfnamefont {M.}~\bibnamefont {Kamionkowski}}\ and\ \bibinfo {author} {\bibfnamefont {A.~G.}\ \bibnamefont {Riess}},\ }\href {\doibase 10.1146/annurev-nucl-111422-024107} {\bibfield  {journal} {\bibinfo  {journal} {Ann. Rev. Nucl. Part. Sci.}\ }\textbf {\bibinfo {volume} {73}},\ \bibinfo {pages} {153} (\bibinfo {year} {2023})},\ \Eprint {http://arxiv.org/abs/2211.04492} {arXiv:2211.04492 [astro-ph.CO]} \BibitemShut {NoStop}%
\bibitem [{\citenamefont {Smith}\ \emph {et~al.}(2020)\citenamefont {Smith}, \citenamefont {Poulin},\ and\ \citenamefont {Amin}}]{Smith2019}%
  \BibitemOpen
  \bibfield  {author} {\bibinfo {author} {\bibfnamefont {T.~L.}\ \bibnamefont {Smith}}, \bibinfo {author} {\bibfnamefont {V.}~\bibnamefont {Poulin}}, \ and\ \bibinfo {author} {\bibfnamefont {M.~A.}\ \bibnamefont {Amin}},\ }\href {\doibase 10.1103/PhysRevD.101.063523} {\bibfield  {journal} {\bibinfo  {journal} {Phys. Rev. D}\ }\textbf {\bibinfo {volume} {101}},\ \bibinfo {pages} {063523} (\bibinfo {year} {2020})},\ \Eprint {http://arxiv.org/abs/1908.06995} {arXiv:1908.06995 [astro-ph.CO]} \BibitemShut {NoStop}%
\bibitem [{\citenamefont {Hill}\ \emph {et~al.}(2020)\citenamefont {Hill}, \citenamefont {McDonough}, \citenamefont {Toomey},\ and\ \citenamefont {Alexander}}]{Hill:2020osr}%
  \BibitemOpen
  \bibfield  {author} {\bibinfo {author} {\bibfnamefont {J.~C.}\ \bibnamefont {Hill}}, \bibinfo {author} {\bibfnamefont {E.}~\bibnamefont {McDonough}}, \bibinfo {author} {\bibfnamefont {M.~W.}\ \bibnamefont {Toomey}}, \ and\ \bibinfo {author} {\bibfnamefont {S.}~\bibnamefont {Alexander}},\ }\href {\doibase 10.1103/PhysRevD.102.043507} {\bibfield  {journal} {\bibinfo  {journal} {Phys. Rev. D}\ }\textbf {\bibinfo {volume} {102}},\ \bibinfo {pages} {043507} (\bibinfo {year} {2020})},\ \Eprint {http://arxiv.org/abs/2003.07355} {arXiv:2003.07355 [astro-ph.CO]} \BibitemShut {NoStop}%
\bibitem [{\citenamefont {Ivanov}\ \emph {et~al.}(2020)\citenamefont {Ivanov}, \citenamefont {McDonough}, \citenamefont {Hill}, \citenamefont {Simonovi\'c}, \citenamefont {Toomey}, \citenamefont {Alexander},\ and\ \citenamefont {Zaldarriaga}}]{Ivanov2020}%
  \BibitemOpen
  \bibfield  {author} {\bibinfo {author} {\bibfnamefont {M.~M.}\ \bibnamefont {Ivanov}}, \bibinfo {author} {\bibfnamefont {E.}~\bibnamefont {McDonough}}, \bibinfo {author} {\bibfnamefont {J.~C.}\ \bibnamefont {Hill}}, \bibinfo {author} {\bibfnamefont {M.}~\bibnamefont {Simonovi\'c}}, \bibinfo {author} {\bibfnamefont {M.~W.}\ \bibnamefont {Toomey}}, \bibinfo {author} {\bibfnamefont {S.}~\bibnamefont {Alexander}}, \ and\ \bibinfo {author} {\bibfnamefont {M.}~\bibnamefont {Zaldarriaga}},\ }\href {\doibase 10.1103/PhysRevD.102.103502} {\bibfield  {journal} {\bibinfo  {journal} {Phys. Rev. D}\ }\textbf {\bibinfo {volume} {102}},\ \bibinfo {pages} {103502} (\bibinfo {year} {2020})},\ \Eprint {http://arxiv.org/abs/2006.11235} {arXiv:2006.11235 [astro-ph.CO]} \BibitemShut {NoStop}%
\bibitem [{\citenamefont {D'Amico}\ \emph {et~al.}(2021)\citenamefont {D'Amico}, \citenamefont {Senatore}, \citenamefont {Zhang},\ and\ \citenamefont {Zheng}}]{DAmico2020}%
  \BibitemOpen
  \bibfield  {author} {\bibinfo {author} {\bibfnamefont {G.}~\bibnamefont {D'Amico}}, \bibinfo {author} {\bibfnamefont {L.}~\bibnamefont {Senatore}}, \bibinfo {author} {\bibfnamefont {P.}~\bibnamefont {Zhang}}, \ and\ \bibinfo {author} {\bibfnamefont {H.}~\bibnamefont {Zheng}},\ }\href {\doibase 10.1088/1475-7516/2021/05/072} {\bibfield  {journal} {\bibinfo  {journal} {JCAP}\ }\textbf {\bibinfo {volume} {05}},\ \bibinfo {pages} {072} (\bibinfo {year} {2021})},\ \Eprint {http://arxiv.org/abs/2006.12420} {arXiv:2006.12420 [astro-ph.CO]} \BibitemShut {NoStop}%
\bibitem [{\citenamefont {La~Posta}\ \emph {et~al.}(2022)\citenamefont {La~Posta}, \citenamefont {Louis}, \citenamefont {Garrido},\ and\ \citenamefont {Hill}}]{LaPosta:2021pgm}%
  \BibitemOpen
  \bibfield  {author} {\bibinfo {author} {\bibfnamefont {A.}~\bibnamefont {La~Posta}}, \bibinfo {author} {\bibfnamefont {T.}~\bibnamefont {Louis}}, \bibinfo {author} {\bibfnamefont {X.}~\bibnamefont {Garrido}}, \ and\ \bibinfo {author} {\bibfnamefont {J.~C.}\ \bibnamefont {Hill}},\ }\href {\doibase 10.1103/PhysRevD.105.083519} {\bibfield  {journal} {\bibinfo  {journal} {Phys. Rev. D}\ }\textbf {\bibinfo {volume} {105}},\ \bibinfo {pages} {083519} (\bibinfo {year} {2022})},\ \Eprint {http://arxiv.org/abs/2112.10754} {arXiv:2112.10754 [astro-ph.CO]} \BibitemShut {NoStop}%
\bibitem [{\citenamefont {Hill}\ \emph {et~al.}(2022)\citenamefont {Hill}, \citenamefont {Calabrese} \emph {et~al.}}]{PhysRevD.105.123536}%
  \BibitemOpen
  \bibfield  {author} {\bibinfo {author} {\bibfnamefont {J.~C.}\ \bibnamefont {Hill}}, \bibinfo {author} {\bibfnamefont {E.}~\bibnamefont {Calabrese}},  \emph {et~al.},\ }\href {\doibase 10.1103/PhysRevD.105.123536} {\bibfield  {journal} {\bibinfo  {journal} {Phys. Rev. D}\ }\textbf {\bibinfo {volume} {105}},\ \bibinfo {pages} {123536} (\bibinfo {year} {2022})}\BibitemShut {NoStop}%
\bibitem [{\citenamefont {Spurio~Mancini}\ \emph {et~al.}(2022)\citenamefont {Spurio~Mancini}, \citenamefont {Piras}, \citenamefont {Alsing}, \citenamefont {Joachimi},\ and\ \citenamefont {Hobson}}]{SpurioMancini:2021ppk}%
  \BibitemOpen
  \bibfield  {author} {\bibinfo {author} {\bibfnamefont {A.}~\bibnamefont {Spurio~Mancini}}, \bibinfo {author} {\bibfnamefont {D.}~\bibnamefont {Piras}}, \bibinfo {author} {\bibfnamefont {J.}~\bibnamefont {Alsing}}, \bibinfo {author} {\bibfnamefont {B.}~\bibnamefont {Joachimi}}, \ and\ \bibinfo {author} {\bibfnamefont {M.~P.}\ \bibnamefont {Hobson}},\ }\href {\doibase 10.1093/mnras/stac064} {\bibfield  {journal} {\bibinfo  {journal} {Mon. Not. Roy. Astron. Soc.}\ }\textbf {\bibinfo {volume} {511}},\ \bibinfo {pages} {1771} (\bibinfo {year} {2022})},\ \Eprint {http://arxiv.org/abs/2106.03846} {arXiv:2106.03846 [astro-ph.CO]} \BibitemShut {NoStop}%
\bibitem [{\citenamefont {Bolliet}\ \emph {et~al.}(2023)\citenamefont {Bolliet}, \citenamefont {Mancini}, \citenamefont {Hill}, \citenamefont {Madhavacheril}, \citenamefont {Jense}, \citenamefont {Calabrese},\ and\ \citenamefont {Dunkley}}]{Bolliet:2023sst}%
  \BibitemOpen
  \bibfield  {author} {\bibinfo {author} {\bibfnamefont {B.}~\bibnamefont {Bolliet}}, \bibinfo {author} {\bibfnamefont {A.~S.}\ \bibnamefont {Mancini}}, \bibinfo {author} {\bibfnamefont {J.~C.}\ \bibnamefont {Hill}}, \bibinfo {author} {\bibfnamefont {M.}~\bibnamefont {Madhavacheril}}, \bibinfo {author} {\bibfnamefont {H.~T.}\ \bibnamefont {Jense}}, \bibinfo {author} {\bibfnamefont {E.}~\bibnamefont {Calabrese}}, \ and\ \bibinfo {author} {\bibfnamefont {J.}~\bibnamefont {Dunkley}},\ }\href@noop {} {\enquote {\bibinfo {title} {High-accuracy emulators for observables in $\lambda$cdm, $n_\mathrm{eff}$, $\sigma m_\nu$, and $w$ cosmologies},}\ } (\bibinfo {year} {2023}),\ \Eprint {http://arxiv.org/abs/2303.01591} {arXiv:2303.01591 [astro-ph.CO]} \BibitemShut {NoStop}%
\bibitem [{\citenamefont {Bolliet}\ \emph {et~al.}(2024)\citenamefont {Bolliet} \emph {et~al.}}]{Bolliet:2023eob}%
  \BibitemOpen
  \bibfield  {author} {\bibinfo {author} {\bibfnamefont {B.}~\bibnamefont {Bolliet}} \emph {et~al.},\ }\href {\doibase 10.1051/epjconf/202429300008} {\bibfield  {journal} {\bibinfo  {journal} {EPJ Web Conf.}\ }\textbf {\bibinfo {volume} {293}},\ \bibinfo {pages} {00008} (\bibinfo {year} {2024})},\ \Eprint {http://arxiv.org/abs/2310.18482} {arXiv:2310.18482 [astro-ph.IM]} \BibitemShut {NoStop}%
\bibitem [{\citenamefont {Torrado}\ and\ \citenamefont {Lewis}(2021{\natexlab{a}})}]{Torrado:2020dgo}%
  \BibitemOpen
  \bibfield  {author} {\bibinfo {author} {\bibfnamefont {J.}~\bibnamefont {Torrado}}\ and\ \bibinfo {author} {\bibfnamefont {A.}~\bibnamefont {Lewis}},\ }\href {\doibase 10.1088/1475-7516/2021/05/057} {\bibfield  {journal} {\bibinfo  {journal} {JCAP}\ }\textbf {\bibinfo {volume} {05}},\ \bibinfo {pages} {057} (\bibinfo {year} {2021}{\natexlab{a}})},\ \Eprint {http://arxiv.org/abs/2005.05290} {arXiv:2005.05290 [astro-ph.IM]} \BibitemShut {NoStop}%
\bibitem [{\citenamefont {Efstathiou}\ \emph {et~al.}(2023)\citenamefont {Efstathiou}, \citenamefont {Rosenberg},\ and\ \citenamefont {Poulin}}]{efstathiou2023improved}%
  \BibitemOpen
  \bibfield  {author} {\bibinfo {author} {\bibfnamefont {G.}~\bibnamefont {Efstathiou}}, \bibinfo {author} {\bibfnamefont {E.}~\bibnamefont {Rosenberg}}, \ and\ \bibinfo {author} {\bibfnamefont {V.}~\bibnamefont {Poulin}},\ }\href@noop {} {\enquote {\bibinfo {title} {Improved planck constraints on axion-like early dark energy as a resolution of the hubble tension},}\ } (\bibinfo {year} {2023}),\ \Eprint {http://arxiv.org/abs/2311.00524} {arXiv:2311.00524 [astro-ph.CO]} \BibitemShut {NoStop}%
\bibitem [{\citenamefont {Akrami}\ \emph {et~al.}(2020)\citenamefont {Akrami} \emph {et~al.}}]{Planck:2020olo}%
  \BibitemOpen
  \bibfield  {author} {\bibinfo {author} {\bibfnamefont {Y.}~\bibnamefont {Akrami}} \emph {et~al.} (\bibinfo {collaboration} {Planck}),\ }\href {\doibase 10.1051/0004-6361/202038073} {\bibfield  {journal} {\bibinfo  {journal} {Astron. Astrophys.}\ }\textbf {\bibinfo {volume} {643}},\ \bibinfo {pages} {A42} (\bibinfo {year} {2020})},\ \Eprint {http://arxiv.org/abs/2007.04997} {arXiv:2007.04997 [astro-ph.CO]} \BibitemShut {NoStop}%
\bibitem [{\citenamefont {Rosenberg}\ \emph {et~al.}(2022)\citenamefont {Rosenberg}, \citenamefont {Gratton},\ and\ \citenamefont {Efstathiou}}]{Rosenberg:2022sdy}%
  \BibitemOpen
  \bibfield  {author} {\bibinfo {author} {\bibfnamefont {E.}~\bibnamefont {Rosenberg}}, \bibinfo {author} {\bibfnamefont {S.}~\bibnamefont {Gratton}}, \ and\ \bibinfo {author} {\bibfnamefont {G.}~\bibnamefont {Efstathiou}},\ }\href {\doibase 10.1093/mnras/stac2744} {\bibfield  {journal} {\bibinfo  {journal} {Mon. Not. Roy. Astron. Soc.}\ }\textbf {\bibinfo {volume} {517}},\ \bibinfo {pages} {4620} (\bibinfo {year} {2022})},\ \Eprint {http://arxiv.org/abs/2205.10869} {arXiv:2205.10869 [astro-ph.CO]} \BibitemShut {NoStop}%
\bibitem [{\citenamefont {{Pagano}}\ \emph {et~al.}(2020)\citenamefont {{Pagano}}, \citenamefont {{Delouis}}, \citenamefont {{Mottet}}, \citenamefont {{Puget}},\ and\ \citenamefont {{Vibert}}}]{pagano/etal:2020}%
  \BibitemOpen
  \bibfield  {author} {\bibinfo {author} {\bibfnamefont {L.}~\bibnamefont {{Pagano}}}, \bibinfo {author} {\bibfnamefont {J.~M.}\ \bibnamefont {{Delouis}}}, \bibinfo {author} {\bibfnamefont {S.}~\bibnamefont {{Mottet}}}, \bibinfo {author} {\bibfnamefont {J.~L.}\ \bibnamefont {{Puget}}}, \ and\ \bibinfo {author} {\bibfnamefont {L.}~\bibnamefont {{Vibert}}},\ }\href {\doibase 10.1051/0004-6361/201936630} {\bibfield  {journal} {\bibinfo  {journal} {\aap}\ }\textbf {\bibinfo {volume} {635}},\ \bibinfo {eid} {A99} (\bibinfo {year} {2020})},\ \Eprint {http://arxiv.org/abs/1908.09856} {arXiv:1908.09856 [astro-ph.CO]} \BibitemShut {NoStop}%
\bibitem [{\citenamefont {Qu}\ \emph {et~al.}(2024)\citenamefont {Qu} \emph {et~al.}}]{ACT:2023dou}%
  \BibitemOpen
  \bibfield  {author} {\bibinfo {author} {\bibfnamefont {F.~J.}\ \bibnamefont {Qu}} \emph {et~al.} (\bibinfo {collaboration} {ACT}),\ }\href {\doibase 10.3847/1538-4357/acfe06} {\bibfield  {journal} {\bibinfo  {journal} {Astrophys. J.}\ }\textbf {\bibinfo {volume} {962}},\ \bibinfo {pages} {112} (\bibinfo {year} {2024})},\ \Eprint {http://arxiv.org/abs/2304.05202} {arXiv:2304.05202 [astro-ph.CO]} \BibitemShut {NoStop}%
\bibitem [{\citenamefont {Madhavacheril}\ \emph {et~al.}(2024)\citenamefont {Madhavacheril} \emph {et~al.}}]{ACT:2023kun}%
  \BibitemOpen
  \bibfield  {author} {\bibinfo {author} {\bibfnamefont {M.~S.}\ \bibnamefont {Madhavacheril}} \emph {et~al.} (\bibinfo {collaboration} {ACT}),\ }\href {\doibase 10.3847/1538-4357/acff5f} {\bibfield  {journal} {\bibinfo  {journal} {Astrophys. J.}\ }\textbf {\bibinfo {volume} {962}},\ \bibinfo {pages} {113} (\bibinfo {year} {2024})},\ \Eprint {http://arxiv.org/abs/2304.05203} {arXiv:2304.05203 [astro-ph.CO]} \BibitemShut {NoStop}%
\bibitem [{\citenamefont {MacCrann}\ \emph {et~al.}(2023)\citenamefont {MacCrann} \emph {et~al.}}]{ACT:2023ubw}%
  \BibitemOpen
  \bibfield  {author} {\bibinfo {author} {\bibfnamefont {N.}~\bibnamefont {MacCrann}} \emph {et~al.} (\bibinfo {collaboration} {ACT}),\ }\href@noop {} {\bibfield  {journal} {\bibinfo  {journal} {ApJ}\ } (\bibinfo {year} {2023})},\ \Eprint {http://arxiv.org/abs/2304.05196} {arXiv:2304.05196 [astro-ph.CO]} \BibitemShut {NoStop}%
\bibitem [{\citenamefont {{Planck Collaboration}}(2020)}]{planck2018lensing}%
  \BibitemOpen
  \bibfield  {author} {\bibinfo {author} {\bibnamefont {{Planck Collaboration}}},\ }\href {\doibase 10.1051/0004-6361/201833886} {\bibfield  {journal} {\bibinfo  {journal} {Astronomy and Astrophysics}\ }\textbf {\bibinfo {volume} {641}},\ \bibinfo {pages} {A8} (\bibinfo {year} {2020})}\BibitemShut {NoStop}%
\bibitem [{\citenamefont {et~al}(2024)}]{desicollaboration2024desi}%
  \BibitemOpen
  \bibfield  {author} {\bibinfo {author} {\bibfnamefont {D.~C.}\ \bibnamefont {et~al}},\ }\href@noop {} {\enquote {\bibinfo {title} {Desi 2024 vi: Cosmological constraints from the measurements of baryon acoustic oscillations},}\ } (\bibinfo {year} {2024}),\ \Eprint {http://arxiv.org/abs/2404.03002} {arXiv:2404.03002 [astro-ph.CO]} \BibitemShut {NoStop}%
\bibitem [{\citenamefont {{Beutler}}\ \emph {et~al.}(2011)\citenamefont {{Beutler}}, \citenamefont {{Blake}}, \citenamefont {{Colless}}, \citenamefont {{Jones}}, \citenamefont {{Staveley-Smith}}, \citenamefont {{Campbell}}, \citenamefont {{Parker}}, \citenamefont {{Saunders}},\ and\ \citenamefont {{Watson}}}]{2011MNRAS.416.3017B}%
  \BibitemOpen
  \bibfield  {author} {\bibinfo {author} {\bibfnamefont {F.}~\bibnamefont {{Beutler}}}, \bibinfo {author} {\bibfnamefont {C.}~\bibnamefont {{Blake}}}, \bibinfo {author} {\bibfnamefont {M.}~\bibnamefont {{Colless}}}, \bibinfo {author} {\bibfnamefont {D.~H.}\ \bibnamefont {{Jones}}}, \bibinfo {author} {\bibfnamefont {L.}~\bibnamefont {{Staveley-Smith}}}, \bibinfo {author} {\bibfnamefont {L.}~\bibnamefont {{Campbell}}}, \bibinfo {author} {\bibfnamefont {Q.}~\bibnamefont {{Parker}}}, \bibinfo {author} {\bibfnamefont {W.}~\bibnamefont {{Saunders}}}, \ and\ \bibinfo {author} {\bibfnamefont {F.}~\bibnamefont {{Watson}}},\ }\href {\doibase 10.1111/j.1365-2966.2011.19250.x} {\bibfield  {journal} {\bibinfo  {journal} {\mnras}\ }\textbf {\bibinfo {volume} {416}},\ \bibinfo {pages} {3017} (\bibinfo {year} {2011})},\ \Eprint {http://arxiv.org/abs/1106.3366} {arXiv:1106.3366 [astro-ph.CO]} \BibitemShut {NoStop}%
\bibitem [{\citenamefont {{Ross}}\ \emph {et~al.}(2015)\citenamefont {{Ross}}, \citenamefont {{Samushia}}, \citenamefont {{Howlett}}, \citenamefont {{Percival}}, \citenamefont {{Burden}},\ and\ \citenamefont {{Manera}}}]{1409.3242}%
  \BibitemOpen
  \bibfield  {author} {\bibinfo {author} {\bibfnamefont {A.~J.}\ \bibnamefont {{Ross}}}, \bibinfo {author} {\bibfnamefont {L.}~\bibnamefont {{Samushia}}}, \bibinfo {author} {\bibfnamefont {C.}~\bibnamefont {{Howlett}}}, \bibinfo {author} {\bibfnamefont {W.~J.}\ \bibnamefont {{Percival}}}, \bibinfo {author} {\bibfnamefont {A.}~\bibnamefont {{Burden}}}, \ and\ \bibinfo {author} {\bibfnamefont {M.}~\bibnamefont {{Manera}}},\ }\href {\doibase 10.1093/mnras/stv154} {\bibfield  {journal} {\bibinfo  {journal} {\mnras}\ }\textbf {\bibinfo {volume} {449}},\ \bibinfo {pages} {835} (\bibinfo {year} {2015})},\ \Eprint {http://arxiv.org/abs/1409.3242} {arXiv:1409.3242 [astro-ph.CO]} \BibitemShut {NoStop}%
\bibitem [{\citenamefont {{Alam}}\ \emph {et~al.}(2017)\citenamefont {{Alam}} \emph {et~al.}}]{1607.03155}%
  \BibitemOpen
  \bibfield  {author} {\bibinfo {author} {\bibfnamefont {S.}~\bibnamefont {{Alam}}} \emph {et~al.},\ }\href {\doibase 10.1093/mnras/stx721} {\bibfield  {journal} {\bibinfo  {journal} {\mnras}\ }\textbf {\bibinfo {volume} {470}},\ \bibinfo {pages} {2617} (\bibinfo {year} {2017})},\ \Eprint {http://arxiv.org/abs/1607.03155} {arXiv:1607.03155 [astro-ph.CO]} \BibitemShut {NoStop}%
\bibitem [{\citenamefont {{Alam}}\ \emph {et~al.}(2021)\citenamefont {{Alam}} \emph {et~al.}}]{2007.08991}%
  \BibitemOpen
  \bibfield  {author} {\bibinfo {author} {\bibfnamefont {S.}~\bibnamefont {{Alam}}} \emph {et~al.},\ }\href {\doibase 10.1103/PhysRevD.103.083533} {\bibfield  {journal} {\bibinfo  {journal} {\prd}\ }\textbf {\bibinfo {volume} {103}},\ \bibinfo {eid} {083533} (\bibinfo {year} {2021})},\ \Eprint {http://arxiv.org/abs/2007.08991} {arXiv:2007.08991 [astro-ph.CO]} \BibitemShut {NoStop}%
\bibitem [{\citenamefont {Scolnic}\ \emph {et~al.}(2022)\citenamefont {Scolnic}, \citenamefont {Brout}, \citenamefont {Carr}, \citenamefont {Riess}, \citenamefont {Davis}, \citenamefont {Dwomoh}, \citenamefont {Jones}, \citenamefont {Ali}, \citenamefont {Charvu}, \citenamefont {Chen}, \citenamefont {Peterson}, \citenamefont {Popovic}, \citenamefont {Rose}, \citenamefont {Wood}, \citenamefont {Brown}, \citenamefont {Chambers}, \citenamefont {Coulter}, \citenamefont {Dettman}, \citenamefont {Dimitriadis}, \citenamefont {Filippenko}, \citenamefont {Foley}, \citenamefont {Jha}, \citenamefont {Kilpatrick}, \citenamefont {Kirshner}, \citenamefont {Pan}, \citenamefont {Rest}, \citenamefont {Rojas-Bravo}, \citenamefont {Siebert}, \citenamefont {Stahl},\ and\ \citenamefont {Zheng}}]{Scolnic_2022}%
  \BibitemOpen
  \bibfield  {author} {\bibinfo {author} {\bibfnamefont {D.}~\bibnamefont {Scolnic}}, \bibinfo {author} {\bibfnamefont {D.}~\bibnamefont {Brout}}, \bibinfo {author} {\bibfnamefont {A.}~\bibnamefont {Carr}}, \bibinfo {author} {\bibfnamefont {A.~G.}\ \bibnamefont {Riess}}, \bibinfo {author} {\bibfnamefont {T.~M.}\ \bibnamefont {Davis}}, \bibinfo {author} {\bibfnamefont {A.}~\bibnamefont {Dwomoh}}, \bibinfo {author} {\bibfnamefont {D.~O.}\ \bibnamefont {Jones}}, \bibinfo {author} {\bibfnamefont {N.}~\bibnamefont {Ali}}, \bibinfo {author} {\bibfnamefont {P.}~\bibnamefont {Charvu}}, \bibinfo {author} {\bibfnamefont {R.}~\bibnamefont {Chen}}, \bibinfo {author} {\bibfnamefont {E.~R.}\ \bibnamefont {Peterson}}, \bibinfo {author} {\bibfnamefont {B.}~\bibnamefont {Popovic}}, \bibinfo {author} {\bibfnamefont {B.~M.}\ \bibnamefont {Rose}}, \bibinfo {author} {\bibfnamefont {C.~M.}\ \bibnamefont {Wood}}, \bibinfo {author} {\bibfnamefont {P.~J.}\ \bibnamefont {Brown}}, \bibinfo {author} {\bibfnamefont {K.}~\bibnamefont
  {Chambers}}, \bibinfo {author} {\bibfnamefont {D.~A.}\ \bibnamefont {Coulter}}, \bibinfo {author} {\bibfnamefont {K.~G.}\ \bibnamefont {Dettman}}, \bibinfo {author} {\bibfnamefont {G.}~\bibnamefont {Dimitriadis}}, \bibinfo {author} {\bibfnamefont {A.~V.}\ \bibnamefont {Filippenko}}, \bibinfo {author} {\bibfnamefont {R.~J.}\ \bibnamefont {Foley}}, \bibinfo {author} {\bibfnamefont {S.~W.}\ \bibnamefont {Jha}}, \bibinfo {author} {\bibfnamefont {C.~D.}\ \bibnamefont {Kilpatrick}}, \bibinfo {author} {\bibfnamefont {R.~P.}\ \bibnamefont {Kirshner}}, \bibinfo {author} {\bibfnamefont {Y.-C.}\ \bibnamefont {Pan}}, \bibinfo {author} {\bibfnamefont {A.}~\bibnamefont {Rest}}, \bibinfo {author} {\bibfnamefont {C.}~\bibnamefont {Rojas-Bravo}}, \bibinfo {author} {\bibfnamefont {M.~R.}\ \bibnamefont {Siebert}}, \bibinfo {author} {\bibfnamefont {B.~E.}\ \bibnamefont {Stahl}}, \ and\ \bibinfo {author} {\bibfnamefont {W.}~\bibnamefont {Zheng}},\ }\href {\doibase 10.3847/1538-4357/ac8b7a} {\bibfield  {journal} {\bibinfo
  {journal} {The Astrophysical Journal}\ }\textbf {\bibinfo {volume} {938}},\ \bibinfo {pages} {113} (\bibinfo {year} {2022})}\BibitemShut {NoStop}%
\bibitem [{\citenamefont {Brout}\ \emph {et~al.}(2022)\citenamefont {Brout} \emph {et~al.}}]{Brout:2022vxf}%
  \BibitemOpen
  \bibfield  {author} {\bibinfo {author} {\bibfnamefont {D.}~\bibnamefont {Brout}} \emph {et~al.},\ }\href {\doibase 10.3847/1538-4357/ac8e04} {\bibfield  {journal} {\bibinfo  {journal} {Astrophys. J.}\ }\textbf {\bibinfo {volume} {938}},\ \bibinfo {pages} {110} (\bibinfo {year} {2022})},\ \Eprint {http://arxiv.org/abs/2202.04077} {arXiv:2202.04077 [astro-ph.CO]} \BibitemShut {NoStop}%
\bibitem [{\citenamefont {Reeves}\ \emph {et~al.}(2023)\citenamefont {Reeves}, \citenamefont {Herold}, \citenamefont {Vagnozzi}, \citenamefont {Sherwin},\ and\ \citenamefont {Ferreira}}]{Reeves_2023}%
  \BibitemOpen
  \bibfield  {author} {\bibinfo {author} {\bibfnamefont {A.}~\bibnamefont {Reeves}}, \bibinfo {author} {\bibfnamefont {L.}~\bibnamefont {Herold}}, \bibinfo {author} {\bibfnamefont {S.}~\bibnamefont {Vagnozzi}}, \bibinfo {author} {\bibfnamefont {B.~D.}\ \bibnamefont {Sherwin}}, \ and\ \bibinfo {author} {\bibfnamefont {E.~G.~M.}\ \bibnamefont {Ferreira}},\ }\href {\doibase 10.1093/mnras/stad317} {\bibfield  {journal} {\bibinfo  {journal} {Monthly Notices of the Royal Astronomical Society}\ }\textbf {\bibinfo {volume} {520}},\ \bibinfo {pages} {3688–3695} (\bibinfo {year} {2023})}\BibitemShut {NoStop}%
\bibitem [{\citenamefont {Poulin}\ \emph {et~al.}(2021)\citenamefont {Poulin}, \citenamefont {Smith},\ and\ \citenamefont {Bartlett}}]{Poulin:2021bjr}%
  \BibitemOpen
  \bibfield  {author} {\bibinfo {author} {\bibfnamefont {V.}~\bibnamefont {Poulin}}, \bibinfo {author} {\bibfnamefont {T.~L.}\ \bibnamefont {Smith}}, \ and\ \bibinfo {author} {\bibfnamefont {A.}~\bibnamefont {Bartlett}},\ }\href {\doibase 10.1103/PhysRevD.104.123550} {\bibfield  {journal} {\bibinfo  {journal} {Phys. Rev. D}\ }\textbf {\bibinfo {volume} {104}},\ \bibinfo {pages} {123550} (\bibinfo {year} {2021})},\ \Eprint {http://arxiv.org/abs/2109.06229} {arXiv:2109.06229 [astro-ph.CO]} \BibitemShut {NoStop}%
\bibitem [{\citenamefont {Smith}\ \emph {et~al.}(2022)\citenamefont {Smith}, \citenamefont {Lucca}, \citenamefont {Poulin}, \citenamefont {Abellan}, \citenamefont {Balkenhol}, \citenamefont {Benabed}, \citenamefont {Galli},\ and\ \citenamefont {Murgia}}]{Smith:2022}%
  \BibitemOpen
  \bibfield  {author} {\bibinfo {author} {\bibfnamefont {T.~L.}\ \bibnamefont {Smith}}, \bibinfo {author} {\bibfnamefont {M.}~\bibnamefont {Lucca}}, \bibinfo {author} {\bibfnamefont {V.}~\bibnamefont {Poulin}}, \bibinfo {author} {\bibfnamefont {G.~F.}\ \bibnamefont {Abellan}}, \bibinfo {author} {\bibfnamefont {L.}~\bibnamefont {Balkenhol}}, \bibinfo {author} {\bibfnamefont {K.}~\bibnamefont {Benabed}}, \bibinfo {author} {\bibfnamefont {S.}~\bibnamefont {Galli}}, \ and\ \bibinfo {author} {\bibfnamefont {R.}~\bibnamefont {Murgia}},\ }\href {\doibase 10.1103/PhysRevD.106.043526} {\bibfield  {journal} {\bibinfo  {journal} {Phys. Rev. D}\ }\textbf {\bibinfo {volume} {106}},\ \bibinfo {pages} {043526} (\bibinfo {year} {2022})},\ \Eprint {http://arxiv.org/abs/2202.09379} {arXiv:2202.09379 [astro-ph.CO]} \BibitemShut {NoStop}%
\bibitem [{\citenamefont {Choi}\ \emph {et~al.}(2020)\citenamefont {Choi} \emph {et~al.}}]{ACT:2020frw}%
  \BibitemOpen
  \bibfield  {author} {\bibinfo {author} {\bibfnamefont {S.~K.}\ \bibnamefont {Choi}} \emph {et~al.} (\bibinfo {collaboration} {ACT}),\ }\href {\doibase 10.1088/1475-7516/2020/12/045} {\bibfield  {journal} {\bibinfo  {journal} {JCAP}\ }\textbf {\bibinfo {volume} {12}},\ \bibinfo {pages} {045} (\bibinfo {year} {2020})},\ \Eprint {http://arxiv.org/abs/2007.07289} {arXiv:2007.07289 [astro-ph.CO]} \BibitemShut {NoStop}%
\bibitem [{\citenamefont {Aiola}\ \emph {et~al.}(2020)\citenamefont {Aiola} \emph {et~al.}}]{ACT:2020gnv}%
  \BibitemOpen
  \bibfield  {author} {\bibinfo {author} {\bibfnamefont {S.}~\bibnamefont {Aiola}} \emph {et~al.} (\bibinfo {collaboration} {ACT}),\ }\href {\doibase 10.1088/1475-7516/2020/12/047} {\bibfield  {journal} {\bibinfo  {journal} {JCAP}\ }\textbf {\bibinfo {volume} {12}},\ \bibinfo {pages} {047} (\bibinfo {year} {2020})},\ \Eprint {http://arxiv.org/abs/2007.07288} {arXiv:2007.07288 [astro-ph.CO]} \BibitemShut {NoStop}%
\bibitem [{\citenamefont {Benson}\ \emph {et~al.}(2014)\citenamefont {Benson} \emph {et~al.}}]{SPT-3G:2014dbx}%
  \BibitemOpen
  \bibfield  {author} {\bibinfo {author} {\bibfnamefont {B.~A.}\ \bibnamefont {Benson}} \emph {et~al.} (\bibinfo {collaboration} {SPT-3G}),\ }\href {\doibase 10.1117/12.2057305} {\bibfield  {journal} {\bibinfo  {journal} {Proc. SPIE Int. Soc. Opt. Eng.}\ }\textbf {\bibinfo {volume} {9153}},\ \bibinfo {pages} {91531P} (\bibinfo {year} {2014})},\ \Eprint {http://arxiv.org/abs/1407.2973} {arXiv:1407.2973 [astro-ph.IM]} \BibitemShut {NoStop}%
\bibitem [{\citenamefont {Herold}\ \emph {et~al.}(2022)\citenamefont {Herold}, \citenamefont {Ferreira},\ and\ \citenamefont {Komatsu}}]{Herold:2021}%
  \BibitemOpen
  \bibfield  {author} {\bibinfo {author} {\bibfnamefont {L.}~\bibnamefont {Herold}}, \bibinfo {author} {\bibfnamefont {E.~G.~M.}\ \bibnamefont {Ferreira}}, \ and\ \bibinfo {author} {\bibfnamefont {E.}~\bibnamefont {Komatsu}},\ }\href {\doibase 10.3847/2041-8213/ac63a3} {\bibfield  {journal} {\bibinfo  {journal} {Astrophys. J. Lett.}\ }\textbf {\bibinfo {volume} {929}},\ \bibinfo {pages} {L16} (\bibinfo {year} {2022})},\ \Eprint {http://arxiv.org/abs/2112.12140} {arXiv:2112.12140 [astro-ph.CO]} \BibitemShut {NoStop}%
\bibitem [{\citenamefont {Nygaard}\ \emph {et~al.}(2023)\citenamefont {Nygaard}, \citenamefont {Holm}, \citenamefont {Hannestad},\ and\ \citenamefont {Tram}}]{Nygaard:2022}%
  \BibitemOpen
  \bibfield  {author} {\bibinfo {author} {\bibfnamefont {A.}~\bibnamefont {Nygaard}}, \bibinfo {author} {\bibfnamefont {E.~B.}\ \bibnamefont {Holm}}, \bibinfo {author} {\bibfnamefont {S.}~\bibnamefont {Hannestad}}, \ and\ \bibinfo {author} {\bibfnamefont {T.}~\bibnamefont {Tram}},\ }\href {\doibase 10.1088/1475-7516/2023/05/025} {\bibfield  {journal} {\bibinfo  {journal} {JCAP}\ }\textbf {\bibinfo {volume} {05}},\ \bibinfo {pages} {025} (\bibinfo {year} {2023})},\ \Eprint {http://arxiv.org/abs/2205.15726} {arXiv:2205.15726 [astro-ph.IM]} \BibitemShut {NoStop}%
\bibitem [{\citenamefont {Lewis}(2019)}]{Lewis:2019xzd}%
  \BibitemOpen
  \bibfield  {author} {\bibinfo {author} {\bibfnamefont {A.}~\bibnamefont {Lewis}},\ }\href {https://getdist.readthedocs.io} {\enquote {\bibinfo {title} {{GetDist: a Python package for analysing Monte Carlo samples}},}\ } (\bibinfo {year} {2019}),\ \Eprint {http://arxiv.org/abs/1910.13970} {arXiv:1910.13970 [astro-ph.IM]} \BibitemShut {NoStop}%
\bibitem [{\citenamefont {{Lesgourgues}}(2011)}]{2011arXiv1104.2932L}%
  \BibitemOpen
  \bibfield  {author} {\bibinfo {author} {\bibfnamefont {J.}~\bibnamefont {{Lesgourgues}}},\ }\href {\doibase 10.48550/arXiv.1104.2932} {\bibfield  {journal} {\bibinfo  {journal} {arXiv e-prints}\ ,\ \bibinfo {eid} {arXiv:1104.2932}} (\bibinfo {year} {2011})},\ \Eprint {http://arxiv.org/abs/1104.2932} {arXiv:1104.2932 [astro-ph.IM]} \BibitemShut {NoStop}%
\bibitem [{\citenamefont {{Blas}}\ \emph {et~al.}(2011)\citenamefont {{Blas}}, \citenamefont {{Lesgourgues}},\ and\ \citenamefont {{Tram}}}]{2011JCAP...07..034B}%
  \BibitemOpen
  \bibfield  {author} {\bibinfo {author} {\bibfnamefont {D.}~\bibnamefont {{Blas}}}, \bibinfo {author} {\bibfnamefont {J.}~\bibnamefont {{Lesgourgues}}}, \ and\ \bibinfo {author} {\bibfnamefont {T.}~\bibnamefont {{Tram}}},\ }\href {\doibase 10.1088/1475-7516/2011/07/034} {\bibfield  {journal} {\bibinfo  {journal} {\jcap}\ }\textbf {\bibinfo {volume} {2011}},\ \bibinfo {eid} {034} (\bibinfo {year} {2011})},\ \Eprint {http://arxiv.org/abs/1104.2933} {arXiv:1104.2933 [astro-ph.CO]} \BibitemShut {NoStop}%
\bibitem [{\citenamefont {McCarthy}\ \emph {et~al.}(2022)\citenamefont {McCarthy}, \citenamefont {Hill},\ and\ \citenamefont {Madhavacheril}}]{McCarthy:2021lfp}%
  \BibitemOpen
  \bibfield  {author} {\bibinfo {author} {\bibfnamefont {F.}~\bibnamefont {McCarthy}}, \bibinfo {author} {\bibfnamefont {J.~C.}\ \bibnamefont {Hill}}, \ and\ \bibinfo {author} {\bibfnamefont {M.~S.}\ \bibnamefont {Madhavacheril}},\ }\href {\doibase 10.1103/PhysRevD.105.023517} {\bibfield  {journal} {\bibinfo  {journal} {Phys. Rev. D}\ }\textbf {\bibinfo {volume} {105}},\ \bibinfo {pages} {023517} (\bibinfo {year} {2022})},\ \Eprint {http://arxiv.org/abs/2103.05582} {arXiv:2103.05582 [astro-ph.CO]} \BibitemShut {NoStop}%
\bibitem [{\citenamefont {Seager}\ \emph {et~al.}(1999)\citenamefont {Seager}, \citenamefont {Sasselov},\ and\ \citenamefont {Scott}}]{Seager:1999bc}%
  \BibitemOpen
  \bibfield  {author} {\bibinfo {author} {\bibfnamefont {S.}~\bibnamefont {Seager}}, \bibinfo {author} {\bibfnamefont {D.~D.}\ \bibnamefont {Sasselov}}, \ and\ \bibinfo {author} {\bibfnamefont {D.}~\bibnamefont {Scott}},\ }\href {\doibase 10.1086/312250} {\bibfield  {journal} {\bibinfo  {journal} {Astrophys. J. Lett.}\ }\textbf {\bibinfo {volume} {523}},\ \bibinfo {pages} {L1} (\bibinfo {year} {1999})},\ \Eprint {http://arxiv.org/abs/astro-ph/9909275} {arXiv:astro-ph/9909275} \BibitemShut {NoStop}%
\bibitem [{\citenamefont {{Scott}}\ and\ \citenamefont {{Moss}}(2009)}]{2009MNRAS.397..445S}%
  \BibitemOpen
  \bibfield  {author} {\bibinfo {author} {\bibfnamefont {D.}~\bibnamefont {{Scott}}}\ and\ \bibinfo {author} {\bibfnamefont {A.}~\bibnamefont {{Moss}}},\ }\href {\doibase 10.1111/j.1365-2966.2009.14939.x} {\bibfield  {journal} {\bibinfo  {journal} {\mnras}\ }\textbf {\bibinfo {volume} {397}},\ \bibinfo {pages} {445} (\bibinfo {year} {2009})},\ \Eprint {http://arxiv.org/abs/0902.3438} {arXiv:0902.3438 [astro-ph.CO]} \BibitemShut {NoStop}%
\bibitem [{\citenamefont {{Chluba}}\ \emph {et~al.}(2012)\citenamefont {{Chluba}}, \citenamefont {{Fung}},\ and\ \citenamefont {{Switzer}}}]{2012MNRAS.423.3227C}%
  \BibitemOpen
  \bibfield  {author} {\bibinfo {author} {\bibfnamefont {J.}~\bibnamefont {{Chluba}}}, \bibinfo {author} {\bibfnamefont {J.}~\bibnamefont {{Fung}}}, \ and\ \bibinfo {author} {\bibfnamefont {E.~R.}\ \bibnamefont {{Switzer}}},\ }\href {\doibase 10.1111/j.1365-2966.2012.21110.x} {\bibfield  {journal} {\bibinfo  {journal} {\mnras}\ }\textbf {\bibinfo {volume} {423}},\ \bibinfo {pages} {3227} (\bibinfo {year} {2012})},\ \Eprint {http://arxiv.org/abs/1110.0247} {arXiv:1110.0247 [astro-ph.CO]} \BibitemShut {NoStop}%
\bibitem [{\citenamefont {Consiglio}\ \emph {et~al.}(2018)\citenamefont {Consiglio}, \citenamefont {de~Salas}, \citenamefont {Mangano}, \citenamefont {Miele}, \citenamefont {Pastor},\ and\ \citenamefont {Pisanti}}]{Consiglio:2017pot}%
  \BibitemOpen
  \bibfield  {author} {\bibinfo {author} {\bibfnamefont {R.}~\bibnamefont {Consiglio}}, \bibinfo {author} {\bibfnamefont {P.~F.}\ \bibnamefont {de~Salas}}, \bibinfo {author} {\bibfnamefont {G.}~\bibnamefont {Mangano}}, \bibinfo {author} {\bibfnamefont {G.}~\bibnamefont {Miele}}, \bibinfo {author} {\bibfnamefont {S.}~\bibnamefont {Pastor}}, \ and\ \bibinfo {author} {\bibfnamefont {O.}~\bibnamefont {Pisanti}},\ }\href {\doibase 10.1016/j.cpc.2018.06.022} {\bibfield  {journal} {\bibinfo  {journal} {Comput. Phys. Commun.}\ }\textbf {\bibinfo {volume} {233}},\ \bibinfo {pages} {237} (\bibinfo {year} {2018})},\ \Eprint {http://arxiv.org/abs/1712.04378} {arXiv:1712.04378 [astro-ph.CO]} \BibitemShut {NoStop}%
\bibitem [{\citenamefont {Mead}\ \emph {et~al.}(2016)\citenamefont {Mead}, \citenamefont {Heymans}, \citenamefont {Lombriser}, \citenamefont {Peacock}, \citenamefont {Steele},\ and\ \citenamefont {Winther}}]{Mead:2016zqy}%
  \BibitemOpen
  \bibfield  {author} {\bibinfo {author} {\bibfnamefont {A.}~\bibnamefont {Mead}}, \bibinfo {author} {\bibfnamefont {C.}~\bibnamefont {Heymans}}, \bibinfo {author} {\bibfnamefont {L.}~\bibnamefont {Lombriser}}, \bibinfo {author} {\bibfnamefont {J.}~\bibnamefont {Peacock}}, \bibinfo {author} {\bibfnamefont {O.}~\bibnamefont {Steele}}, \ and\ \bibinfo {author} {\bibfnamefont {H.}~\bibnamefont {Winther}},\ }\href {\doibase 10.1093/mnras/stw681} {\bibfield  {journal} {\bibinfo  {journal} {Mon. Not. Roy. Astron. Soc.}\ }\textbf {\bibinfo {volume} {459}},\ \bibinfo {pages} {1468} (\bibinfo {year} {2016})},\ \Eprint {http://arxiv.org/abs/1602.02154} {arXiv:1602.02154 [astro-ph.CO]} \BibitemShut {NoStop}%
\bibitem [{\citenamefont {Torrado}\ and\ \citenamefont {Lewis}(2021{\natexlab{b}})}]{cobaya}%
  \BibitemOpen
  \bibfield  {author} {\bibinfo {author} {\bibfnamefont {J.}~\bibnamefont {Torrado}}\ and\ \bibinfo {author} {\bibfnamefont {A.}~\bibnamefont {Lewis}},\ }\href {\doibase 10.1088/1475-7516/2021/05/057} {\bibfield  {journal} {\bibinfo  {journal} {Journal of Cosmology and Astroparticle Physics}\ }\textbf {\bibinfo {volume} {2021}},\ \bibinfo {pages} {057} (\bibinfo {year} {2021}{\natexlab{b}})}\BibitemShut {NoStop}%
\end{thebibliography}%

\end{document}